\documentclass{article}

\usepackage{microtype}
\usepackage{graphicx}
\usepackage{subfigure}
\usepackage{booktabs} 

\usepackage{hyperref}

\usepackage[accepted]{icml2022}

\usepackage{amsmath}
\usepackage{amssymb}
\usepackage{mathtools}
\usepackage{amsthm}

\usepackage[capitalize,noabbrev]{cleveref}

\theoremstyle{plain}
\newtheorem{theorem}{Theorem}[section]

\theoremstyle{definition}

\theoremstyle{remark}
\newtheorem{remark}[theorem]{Remark}

\usepackage[textsize=tiny]{todonotes}

\usepackage{xargs}
\usepackage{etoolbox}
\usepackage{ifthen}
\usepackage{tikz}
\usetikzlibrary{calc, fit}
\usepackage{enumitem}
\usepackage{stfloats}

\newenvironment{rmklist}{\begin{enumerate}[label={(\alph*)},topsep=0em,itemsep=0em,itemindent=2em,leftmargin=0em]}{\end{enumerate}}
\newenvironment{romanlist}{\begin{enumerate}[label={(\roman*)},itemsep=0em,itemindent=2em,leftmargin=0em]}{\end{enumerate}}

\newcommand{\vsp}[1]{\vskip #1in}

\newcommand{\opleft}[1]{\mathopen{}\left#1}
\newcommand{\opright}[1]{\right#1\mathclose{}}
\newcommandx{\braces}[4]{%
\ifstrequal{#3}{normal}{#1#4#2}{%
\ifstrequal{#3}{auto}{\left#1#4\right#2}{%
\ifstrequal{#3}{opauto}{\opleft#1#4\opright#2}{%
#3#1#4#3#2}}}%
}

\newcommand{\N}{\mathbb{N}} 
\newcommand{\R}{\mathbb{R}} 

\newcommand{\suchthat}[1][normal]{\ifstrequal{#1}{normal}{\mid}{#1|}} 

\newcommandx{\intvcl}[3][1=normal]{\braces{[}{]}{#1}{#2, #3}} 
\newcommandx{\intvop}[3][1=normal]{\braces{(}{)}{#1}{#2, #3}} 
\newcommandx{\intvclop}[3][1=normal]{\braces{[}{)}{#1}{#2, #3}} 
\newcommandx{\intvopcl}[3][1=normal]{\braces{(}{]}{#1}{#2, #3}} 

\newcommandx{\abs}[2][1=normal]{\braces{\lvert}{\rvert}{#1}{#2}} 
\newcommandx{\ceil}[2][1=normal]{\braces{\lceil}{\rceil}{#1}{#2}} 
\newcommandx{\floor}[2][1=normal]{\braces{\lfloor}{\rfloor}{#1}{#2}} 
\newcommandx{\round}[2][1=normal]{\braces{[}{]}{#1}{#2}} 
\newcommandx{\der}[1]{D^{#1}} 
\newcommandx{\gradient}{\nabla} 
\newcommandx{\partder}[4][1={},4={}]{\frac{\partial^{#4} #2}{\partial #3^{#4}}\ifargdef{#1}{\Big|_{#1}}} 
\newcommandx{\integ}[4][1={},2={}]{\int_{#1}^{#2} #3 \, #4} 
\newcommandx{\asympffaster}[2][1=normal]{o\braces{(}{)}{#1}{#2}} 
\newcommandx{\asympfaster}[2][1=normal]{O\braces{(}{)}{#1}{#2}} 
\newcommandx{\asympeq}[2][1=normal]{\Theta\braces{(}{)}{#1}{#2}} 
\newcommandx{\asympsslower}[2][1=normal]{\omega\braces{(}{)}{#1}{#2}} 
\newcommandx{\asympslower}[2][1=normal]{\Omega\braces{(}{)}{#1}{#2}} 

\newcommandx{\norm}[2][1=normal]{\braces{\|}{\|}{#1}{#2}} 
\renewcommandx{\sp}[3][1=normal]{\braces{\langle}{\rangle}{#1}{#2, #3}} 
\newcommandx{\End}[2][2={}]{\mathcal{L}\opleft( #1 \ifargdef{#2}{, #2} \opright)} 
\newcommand{\T}{\mathsf{T}} 
\renewcommand{\vec}[1]{\boldsymbol{#1}} 
\newcommandx{\opnorm}[2][1=normal]{\norm[#1]{#2}_{\operatorname{op}}} 
\newcommandx{\ball}[2][1={},2={}]{B_{#1}^{#2}} 

\newcommandx{\measure}[2][1=normal]{\operatorname{vol}\braces{(}{)}{#1}{#2}} 
\newcommandx{\Leb}[3][1={},3=normal]{L^{#2}\ifargdef{#1}{\braces{(}{)}{#3}{#1}}{}} 
\newcommandx{\Lebnorm}[4][1=normal,3={2},4={}]{\norm[#1]{#2}_{#3}} 
\renewcommandx{\l}[3][1={},3=normal]{\ell_{#2}\ifargdef{#1}{\braces{(}{)}{#3}{#1}}} 
\newcommandx{\lnorm}[4][1=normal,3={2},4={}]{\norm[#1]{#2}_{#3}} 
\newcommandx{\Smooth}[4][1={},3={},4=normal]{C_{#3}^{#2}\ifargdef{#1}{\braces{(}{)}{#4}{#1}}} 
\newcommandx{\Schwartz}[2][1={},2=normal]{\mathscr{S}\ifargdef{#1}{\braces{(}{)}{#2}{#1}}} 
\newcommandx{\Schwartzpoly}[2][1=normal]{\braces{\langle}{\rangle}{#1}{\abs[#1]{#2}} } 
\newcommandx{\Tempdistr}[2][1={},2=normal]{\mathscr{S}'\ifargdef{#1}{\braces{(}{)}{#2}{#1}}} 
\newcommandx{\distrinp}[3][1=normal]{\braces{\langle}{\rangle}{#1}{#2, #3}} 
\newcommandx{\ft}[3][1=default,2=auto]{
\ifstrequal{#1}{default}{\widehat{#3}}{
\ifstrequal{#1}{long}{{\braces{(}{)}{#2}{#3}}^{\wedge}}{}}} 
\newcommandx{\ift}[3][1=default,2=auto]{
\ifstrequal{#1}{default}{\check{#3}}{
\ifstrequal{#1}{long}{{\braces{(}{)}{#2}{#3}}^{\vee}}{}}} 

\newcommandx{\prob}[2][1={},2=normal]{\mathbb{P}\ifargdef{#1}{\braces{[}{]}{#2}{#1}}}
\newcommandx{\mean}[2][1={},2=normal]{\mathbb{E}\ifargdef{#1}{\braces{[}{]}{#2}{#1}}}
\newcommandx{\var}[2][1={},2=normal]{\mathbb{V}\ifargdef{#1}{\braces{[}{]}{#2}{#1}}}

\newcommandx{\Unif}[2][1=normal]{\mathcal{U}\braces{(}{)}{#1}{#2}} 
\newcommandx{\Normdistr}[3][1=normal]{\mathcal{N}\braces{(}{)}{#1}{#2, #3}} 
\newcommandx{\Poi}[2][1=normal]{\mathrm{Poi}\braces{(}{)}{#1}{#2}} 
\newcommandx{\normsubg}[2][1=normal]{\norm[#1]{#2}_{\psi_2}} 

\newcommand{\y}{\vec{y}} 

\newcommand{\Noise}{\vec{e}} 

\newcommand{\x}{\vec{x}} 
\newcommand{\xsolu}{\hat{\vec{x}}} 

\newcommand{\DC}{\mathcal{DC}}
\newcommand{\dcparam}{\lambda}
\newcommand{\DCparam}{\vec{\lambda}}
\newcommand{\NNparam}{\vec{\theta}}
\newcommand{\Fanparam}{\vec{\theta}_{\smash{\texttt{fan}}}}
\newcommand{\AFan}{\vec{F}} 
\newcommand{\FBP}{\texttt{FBP}} 

\newcommand{\UNet}{\textrm{UNet}}

\newcommand{\Rec}{\mathcal{R}}

\newcommand{\Tira}{\textrm{Tiramisu}}

\newcommand{\ItNet}{\textrm{ItNet}}
\newcommand{\ItNett}{\textrm{ItNet-post}}

\icmltitlerunning{Near-Exact Recovery for Tomographic Inverse Problems via Deep Learning}

\begin{document}

\twocolumn[
\icmltitle{Near-Exact Recovery for Tomographic Inverse Problems via Deep Learning}



\icmlsetsymbol{equal}{*}

\begin{icmlauthorlist}
\icmlauthor{Martin Genzel}{equal,hzb}
\icmlauthor{Ingo G{\"u}hring}{equal,tuberlin}
\icmlauthor{Jan Macdonald}{equal,tuberlin}
\icmlauthor{Maximilian M{\"a}rz}{equal,tuberlin}
\end{icmlauthorlist}

\icmlaffiliation{hzb}{Helmholtz-Zentrum Berlin f{\"u}r Materialien und Energie, Germany (work done while at Utrecht University, Netherlands)}
\icmlaffiliation{tuberlin}{Technical University Berlin, Germany}

\icmlcorrespondingauthor{Martin Genzel}{martingenzel@gmail.com}

\icmlkeywords{Inverse problems, computed tomography, computational imaging, deep learning, end-to-end neural networks}

\vskip 0.3in
]



\printAffiliationsAndNotice{\icmlEqualContribution (the authors are ordered alphabetically by last name).} 

\begin{abstract}
This work is concerned with the following fundamental question in scientific machine learning: Can deep-learning-based methods solve noise-free inverse problems to near-perfect accuracy?
Positive evidence is provided for the first time, focusing on a prototypical computed tomography (CT) setup.
We demonstrate that an iterative end-to-end network scheme enables reconstructions close to numerical precision, comparable to classical compressed sensing strategies.
Our results build on our winning submission to the recent AAPM DL-Sparse-View CT Challenge.
Its goal was to identify the state-of-the-art in solving the sparse-view CT inverse problem with data-driven techniques.
A specific difficulty of the challenge setup was that the precise forward model remained unknown to the participants.
Therefore, a key feature of our approach was to initially estimate the unknown fanbeam geometry in a data-driven calibration step.
Apart from an in-depth analysis of our methodology, we also demonstrate its state-of-the-art performance on the open-access real-world dataset LoDoPaB CT.
\end{abstract}

\section{Introduction}
\label{sec:introduction}

In recent years, deep learning methods have been successfully applied to many problems of the natural sciences \cite{lbh15,sch15,gbc16}.
A prominent example of such \emph{scientific machine learning} is the development of efficient solutions strategies for inverse problems \cite{amos19,ojbmdw20}, such as those encountered in medical imaging.
But despite unprecedented empirical performance in various practical scenarios, a lack of evidence for the reliability of these methods remains. 
For instance, \citet{sidky20} have recently demonstrated that \emph{post-processing} of filtered backprojection images with the prominent UNet-architecture may not yield satisfactory recovery precision in sparse-view computed tomography (CT).
This observation gave rise to the recent AAPM Grand Challenge ``Deep Learning for Inverse Problems: Sparse-View Computed Tomography Image Reconstruction'', with the goal \emph{``to identify the state-of-the-art in solving the CT inverse problem with data-driven techniques''}~\cite{sid+21}.
Here, the term `solving' is used to describe algorithms that provide perfect recovery from incomplete, noiseless measurements.

The study of this desirable property was popularized by the field of \emph{compressed sensing} \cite{crt06a,don06,fh13}.
Indeed, high precision in the noiseless, undersampled regime can be used to benchmark reconstruction methods and is a driving factor for their acceptance in practice.
Therefore, an important open research question is whether deep-learning-based schemes can achieve such \mbox{(near-)perfect} solutions, comparable to model-based algorithms like total variation (TV) minimization.
The present article makes first progress in this direction, building on our winning submission to the AAPM challenge. Our main contributions are as follows:
\begin{romanlist}
	\item
	We show that end-to-end neural networks can achieve \emph{near-perfect accuracy} on the prescribed CT reconstruction task.
	This underscores the reliability of deep-learning-based solvers for inverse problems, in the sense that they can match the precision of a widely-accepted benchmark (TV minimization) in the noiseless limit.
	\item
	We give a \emph{detailed analysis} of our solution strategy, which has significantly outperformed the runner-up teams.
	Although the challenge amounts to a comparison with 24 competing methods, we also explicitly demonstrate the superiority over several popular baselines in this work.
	This includes the learned primal-dual algorithm~\cite{ao18}, which is commonly considered as state-of-the-art for solving inverse problems, cf.~\citet{leuschner21,ramzi20}.
	In addition, we show the effectiveness of our learning pipeline beyond synthetically generated image data:
	The proposed neural network scheme produces state-of-the-art results on the LoDoPaB CT dataset~\cite{leuschner21}, currently ranked first in the public leaderboard.
	\item
	We distill several insights of broader interest and conceptual value.
	Most notably, we found that \emph{simple building blocks} (e.g., end-to-end training, alternation between learned and model-based components, etc.) and a careful \emph{pre-training strategy} already allow for remarkable performance gains.
	\item
	A specific difficulty of the AAPM challenge setup was an \emph{unknown (fanbeam) forward model}.
	Therefore, a crucial step of our approach consists in the data-driven estimation of the underlying fanbeam geometry.
	We accomplish this by fitting a generic, parameterized fanbeam operator to the provided sinogram-image pairs in a deep-learning-like fashion (i.e., by gradient descent with backpropagation/automatic differentiation). This conception may find further application in the context of geometric calibration and forward operator correction.
\end{romanlist}

Conceptually, our approach stems from the following (debatable) observation:
\begin{quote}\itshape
	High reconstruction accuracy is only possible if the forward model is explicitly incorporated into the solution map, e.g., by an iterative promotion of data-consistency.
\end{quote}
The vital role of the forward operator in data-driven solutions to inverse problems is by no means a new insight.
It is well in line with a central pillar of scientific machine learning, namely that neural networks can be often enriched (or constrained) by physical modeling.
Indeed, the seminal works on deep learning techniques for inverse problems are inspired by \emph{unrolling} classical algorithms, e.g., see~\citet{kl10,yslx16,ham+17,amj18,ao18,che+18,sch+19,hsqdsr19,chlf+20,hfgy21,gow21}.
At the present time, most state-of-the-art methods rely on \emph{iterative} end-to-end networks and related schemes, e.g., see~\citet{kno+20b,muc+20,leuschner21} for other recent competition benchmarks.

Our contribution to the AAPM challenge is no exception in that respect.
We propose a conceptually simple, yet powerful deep learning pipeline, which turns a post-processing UNet~\cite{rfb15} into an iterative reconstruction scheme.
While many of its individual components have been previously reported in the literature, the overall strategy is novel.
Our design differs from more common unrolled networks in several aspects, most notably the following two:
(a)~we make use of a \emph{pre-trained} UNet as the computational backbone, and (b)~data-consistency is inspired by an \mbox{$\ell^2$-gradient} step, but employs the \emph{filtered} backprojection (FBP) instead of the regular adjoint.
In line with most previous works, our unrolled network only involves very few (five) iteration steps.
However, we are the first to show that this is sufficient to match the precision of model-based solvers, which typically need hundreds or thousands of iterations before convergence (and therefore require significantly more computation time).

\subsection*{The Bigger Picture}

To put our results into a broader context, it is worth considering an inverse problem in its prototypical, finite-dimensional form:
\begin{equation*}
	\y = \AFan \x + \Noise,
\end{equation*}
where $\x \in \R^N$ denotes the unknown (image) signal, $\AFan \in \R^{m \times N}$ the forward operator, and $\y$ are noisy measurements. The goal is to reconstruct $\x$ from $\y$.
The error of any given reconstruction map $\Rec :\nobreak \R^m \to\nobreak \R^N$ can be decomposed as
\begin{equation*}
	\norm[\big]{\x - \Rec(\y)}_2 \leq \underbrace{\norm[\big]{\x - \Rec(\AFan\x)}_2}_{\text{(a)}} + \underbrace{\norm[\big]{\Rec(\AFan\x) - \Rec(\y)}_2}_{\text{(b)}}.
\end{equation*}
The first term (a) is associated with the \emph{accuracy} (or \emph{precision}) of $\Rec$ and measures how well $\x$ can be estimated in the idealistic situation of noiseless measurements.
The second term (b) captures the \emph{robustness} of $\Rec$ against perturbations~$\Noise$ of the measurements.
Adequate control over both expressions forms the backbone of inverse problem theory and scientific computing in general.

The present paper is primarily concerned with the accuracy term~(a), more specifically, the root-mean-square-error (RMSE).
We demonstrate that it can become sufficiently close to zero (on the image data distribution) when $\Rec$ corresponds to a fully data-driven solver.
The competitive setting of the AAPM challenge has provided us with the right benchmark to conduct such a case study.

Regarding the robustness term (b), we refer to the recent work by \citet{genzel20} for an in-depth case study.
It was shown that even standard end-to-end networks can be surprisingly robust against adversarial perturbations (i.e., worst-case noise), comparable to a provably stable benchmark methods.
Together with \citet{genzel20}, the present work provides further evidence for the reliability of deep-learning-based solutions to inverse problems.

\subsection*{Organization of This Article}

The rest of this article is organized as follows.
Section~\ref{sec:challenge_setup} gives a brief overview of the AAPM challenge setup and the training/test-data.
Section~\ref{sec:methodology} is then devoted to a conceptual description of our learning pipeline, while more details on the implementation can be found in Appendix~\ref{sec:exact_challenge_setup}.
Our results and several accompanying experiments are reported in Section~\ref{sec:results_analysis}.
We conclude with a short discussion in Section~\ref{sec:discussion}.
For an evaluation of our method on the LoDoPaB CT dataset, see Appendix~\ref{sec:lodopab}.

\section{AAPM Challenge Setup} \label{sec:challenge_setup}

\begin{figure}[!t]
	\centering
	\includegraphics[width=\columnwidth]{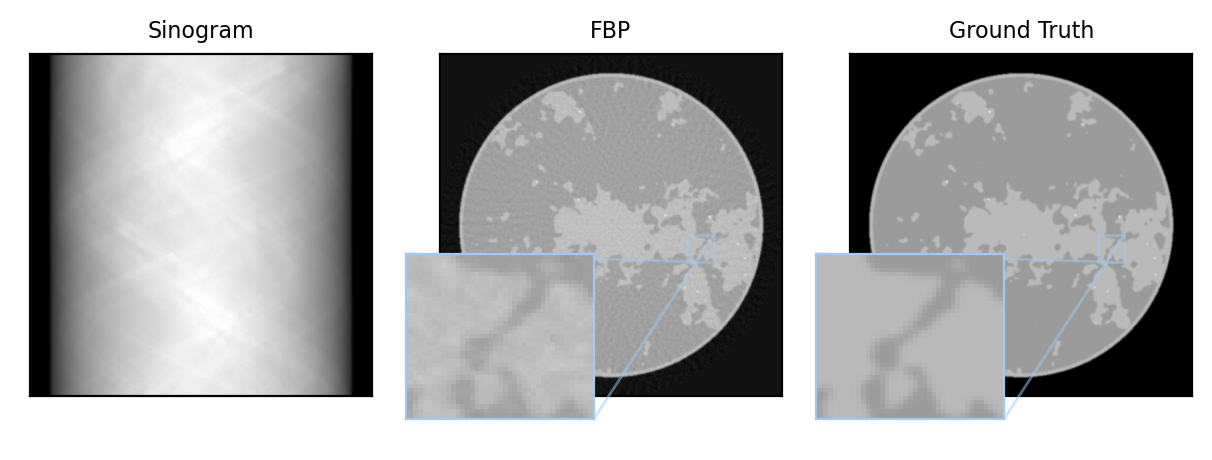}
	\vsp{-.2}
	\caption{\unboldmath\textbf{AAPM challenge data.} Example of a 128-view sinogram, FBP reconstruction, and ground-truth phantom image taken from the AAPM challenge training dataset.}
	\label{fig:data_example}
\end{figure}

The AAPM challenge data is similar to the setting of~\citet{sidky20}, i.e., it is based on synthetic 2D grayscale images of size $512 \times 512$ simulating real-world mid-plane breast CT device scans.
Four different tissues were modeled: adipose, skin, fibroglandular tissue, and microcalcifications.
To obtain smooth transitions at tissue boundaries, Gaussian smoothing was applied.
A fanbeam geometry with 128 projections over 360 degrees was used to create sinograms and FBPs, see Fig.~\ref{fig:data_example} for an example.
Notably, the exact fanbeam geometry was not revealed to the participants.
No noise was added to the data, neither to the phantom images nor the measurements.
The provided training set consisted of 4000 tuples of phantom images, their corresponding 128-view sinograms, and FBP reconstructions.
A test set of 100 pairs of sinograms and FBPs (without publicly available ground-truth phantoms) was used for the final challenge evaluation.

Initially, about 50 international teams have participated, out of which 25 have submitted their method to the final evaluation.
More details about the challenge setup and results can be found in the official challenge report \cite{sp21}.

\section{Methodology}
\label{sec:methodology}

This section gives an overview of our (three-step) methodology for the AAPM challenge and motivates our design choices.\footnote{Our code is available at \url{https://github.com/jmaces/aapm-ct-challenge}.}

\subsection*{Step 1: Data-Driven Geometry Identification}

\begin{figure}[t!]
	\centering
	\begin{tikzpicture}[scale=0.65]

\draw[dashed] (0.5, 0.5) circle[radius=3.0];

\draw[black, ultra thick] (-4.25,-4) -- (5.25,-4) node[below left]{detector array};

\draw[stealth-stealth, black] (-4,-4.25) -- (-2.5,-4.25) node[below]{$s_{\text{detector}}$};

\draw[<-,domain=10:55,thick] plot ({0.5+3.25*cos(\x)}, {0.5+3.25*sin(\x)});
\draw (3.25,2.3) node[rotate=-60,above,scale=0.9,thick] {rotate};

\draw[fill=black, fill opacity=0.1,scale=0.85]  plot[smooth, tension=.7] coordinates {(-1.2,0.5) (-0.8,1.2) (0.2,1.5) (0.7,1) (1.2,1.3) (1.7,1.8) (2.2,0.5) (2.5, 0) (1.9,-0.5) (1.0,-1) (0.2,-0.7) (-0.3,-1.2) (-0.8,-0.8) (-1.25, 0) (-1.2,0.5)};

\draw (0.5,3.5) -- ++(-90:1.5) arc (-90:-121:1.5) node[midway,above]{$\gamma$} -- cycle;


\draw[-stealth,gray] (0.5,3.5) -- (5,-4) node[above right]{\footnotesize $n_{\text{detector}}$};

\draw[-stealth,gray] (0.5,3.5) -- (3.5,-4);

\draw[-stealth,gray] (0.5,3.5) -- (2,-4);

\draw[thick, stealth-stealth, black] (0.5,3.5) -- (0.5,0.5);
\draw (1.25,2) node[black] {$d_{\text{source}}$};

\draw[thick, stealth-stealth, black] (0.5,0.5) -- (0.5,-4);
\draw (1.4,-3) node[black] {$d_{\text{detector}}$};

\draw[-stealth,gray] (0.5,3.5) -- (-1.0,-4); 

\draw[-stealth,gray] (0.5,3.5) -- (-2.5,-4) node[above left]{\footnotesize 2};

\draw[-stealth,gray] (0.5,3.5) -- (-4.0,-4) node[above left]{\footnotesize 1};

\fill[thick, black] (0.5,0.5) circle (2pt);

\fill[thick, black] (0.5,3.5) circle (3pt) node[above] {X-ray source};
\fill[thick, gray] (1.58,3.29) circle (3pt) node[below]{\footnotesize 2};
\fill[thick, gray] (2.59,2.65) circle (3pt) node[below]{\footnotesize 3};
\fill[thick, gray] (-0.74,3.22) circle (3pt) node[above left]{\footnotesize $n_\text{angle}$};
\end{tikzpicture}
	\caption{\unboldmath\textbf{Fanbeam geometry.} Illustration of the parameters determining the geometry of the fanbeam CT model.}
	\label{fig:radon}
\end{figure}
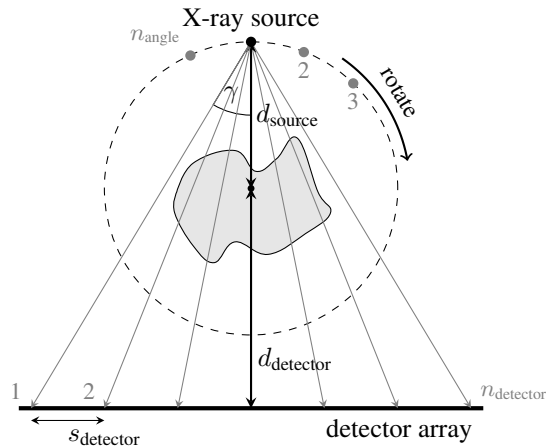

In the first step of our reconstruction pipeline, we estimate the unknown forward operator from the provided training data.
The continuous version of \emph{tomographic fanbeam measurements} is based on computing line integrals:
\begin{equation*}
	p(s,\varphi) = \int_{L(s,\varphi)} x_0(x,y) \, \mathrm{d}(x,y),
\end{equation*}
where $x_0$ is the unknown image and $L(s,\varphi)$ denotes a line in fanbeam coordinates, i.e., $\varphi$ is the \emph{fan rotation angle} and~$s$ encodes the \emph{sensor position}; see \citet{fessler17} for more details.
In an idealized situation, the fanbeam model is specified by the following geometric parameters\footnote{We have found that this basic model was enough to accurately describe the AAPM challenge setup. If needed, it would be possible to account for other factors such as non-flat detector arrays, offsets of the axis of rotation from the origin, misalignments of the detector array, etc.} (see Fig.~\ref{fig:radon} for an illustration):
\begin{itemize}[itemsep=.25em,parsep=0em,topsep=0em]
	\item $d_\text{source}$ -- distance of the X-ray source to the origin,
	\item $d_\text{detector}$ -- distance of the detector array to the origin,
	\item $n_\text{detector}$ -- number of detector elements,
	\item $s_\text{detector}$ -- spacing of detector elements along the array,
	\item $n_\text{angle}$ -- number of fan rotation angles,
	\item $\vec{\varphi}\in [0,2\pi]^{n_\text{angle}}$ -- discrete list of rotation angles.
\end{itemize}
Here, it is assumed that integrals are only measured along a finite number of lines, determined by $m \coloneqq n_\text{detector} \cdot\nobreak n_\text{angle}$.
In the \emph{sparse-view} challenge setup, the resulting forward operator is \emph{severely ill-posed}, since only the measurements of a few fan rotation angles $n_\text{angle}$ are acquired.
Furthermore, the geometric setup is not disclosed to the challenge participants---it is only known that fanbeam measurements are used.

We have addressed this lack of information by a data-driven estimation strategy that fits the above set of parameters to the given training data.
To this end, we first observe that the previous parametrization is redundant, and without of loss of generality, we may assume that $s_\text{detector}=1$ (by rescaling $d_\text{detector}$ appropriately).
Further, if the field-of-view angle $\gamma$ is known, then the relation
\begin{equation}
	\label{eq:simp1}
	d_\text{detector} = \frac{n_\text{detector} \cdot s_\text{detector}}{2\tan{\gamma}}-d_\text{source}
\end{equation}
can be used to eliminate another parameter.
Thus, the fanbeam geometry is effectively determined by the reduced parameter set $(d_\text{source}, n_\text{detector}, n_\text{angle}, \vec{\varphi})$.
The training data provides pairs of discrete images $\x \in\R^{\smash{512\cdot 512 \eqqcolon N}}$ and its simulated fanbeam measurements $\y \in\R^{\smash{128\cdot1024 = m}}$, from which the dimensions $n_{\smash{\text{angle}}}=128$ and $n_{\smash{\text{detector}}} = 1024$ can be derived.
We determine the field of view as $\gamma = \arcsin(256 / d_\text{source})$, so that the  maximum inscribed circle in the discrete image is exactly contained within each fan of lines, which is a common choice for fanbeam CT.
Hence, \eqref{eq:simp1} leads to
\begin{equation*}
	d_\text{detector} = 2 \cdot s_\text{detector} \cdot \sqrt{d_\text{source}^2  - 256^2} - d_\text{source} \ .
\end{equation*}

The main difficulty of Step~1 lies in the estimation of the remaining parameters $(d_\text{source},\vec{\varphi})$.
To that end, we have implemented a discrete fanbeam transform from scratch in \texttt{PyTorch} (together with its corresponding FBP).
A distinctive aspect of our implementation is the use of a vectorized numerical integration that enables the efficient computation of derivatives with respect to the geometric parameters by means of \emph{automatic differentiation}.
This feature can be exploited for a data-driven parameter identification, for instance, by a gradient descent.
More precisely, we use a ray-driven numerical integration for the forward model and a pixel-driven and sinogram-reweighting-based FBP (with a Hamming filter), see \citet[Sec. 3.9.2]{fessler17}.
In addition to the parameters $(d_\text{source},\vec{\varphi})$, we also introduce learnable scaling factors $s_\text{fwd}$ and $s_\text{fbp}$ for the forward and inverse transform, respectively.
They account for ambiguities in chosing the discretization units of distance compared to the actual physical units of distance.

As previously indicated, we estimate the free parameters $\Fanparam = (s_{\smash{\text{fwd}}},d_{\smash{\text{source}}},\vec{\varphi})\in\R^{130}$ of the implemented forward operator $\AFan[\Fanparam]\in\R^{m\times N}$ in a deep-learning-like fashion:
The ability to compute derivatives $\tfrac{\mathrm{d}\AFan}{\mathrm{d}\Fanparam}$ allows us to make use of the $M=4000$ sinogram-image pairs $\{(\y^i, \x^i) \}_{i=1}^M$ by solving
\begin{equation}\label{eq:op_ident}
	\min_{\Fanparam} \ \tfrac{1}{M}\sum_{i=1}^M \norm[\big]{\AFan[\Fanparam] (\x^i) - \y^i}_2^2 
\end{equation}
with a variant of gradient descent (see Remark~\ref{rmk:rmk1} below for details).
Finally, we determine $s_\text{fbp}$ by solving
\begin{equation}\label{eq:fpb_ident}
	\min_{s_\text{fbp}} \ \tfrac{1}{M}\sum_{i=1}^M \norm[\big]{\x^i - \FBP[\Fanparam,s_\text{fbp}](\y^i)}_2^2 \ , 
\end{equation}
while keeping the already identified parameters fixed.
We will use the short-hand notation $\AFan$ and $\FBP$ for the estimated operators $\AFan[\Fanparam]$ and $\FBP[\Fanparam,s_\text{fbp}] \colon \R^m \to \R^N$, respectively.

\begin{remark}
	\label{rmk:rmk1}
	\begin{rmklist}
		\item
		Clearly, the formulation~\eqref{eq:op_ident} is non-convex and therefore it is not clear whether gradient descent enables an accurate estimation of the underlying fanbeam geometry.
		Indeed, standard gradient descent was found to be very sensitive to the initialization of $\Fanparam$ and got stuck in bad local minima.
		To overcome this issue, we solve~\eqref{eq:op_ident} by a \emph{coordinate descent} instead, which alternatingly optimizes over $s_\text{fwd}$, $d_\text{source}$, and $\vec{\varphi}$ with individual learning rates.
		This strategy was found to effectively account for large deviations of gradient magnitudes of the different parameters.
		Indeed, we observed a fast convergence and a reliable identification of $\Fanparam$, independently of the initialization.
		\item In principle, the strategy of~\eqref{eq:op_ident} requires only few training samples to be successful.
		However, when verifying the robustness of the outlined strategy against measurement noise, we observed that it is beneficial to employ more training data.
		\item Subsequent to the estimation of an accurate fanbeam geometry, we still recognized a small systematic error in our forward model.
		We suspect that it is caused by subtle differences in the numerical integration in comparison to the true forward model of the AAPM challenge.
		In compensation, we compute the (pixelwise) mean error over the training set, as an additive correction of the model bias.
	\end{rmklist}
\end{remark}
\subsection*{Step 2: Pre-Training a UNet as Computational Backbone}

The centerpiece of our reconstruction scheme is formed by a standard \emph{UNet-architecture} $\UNet[\NNparam] \colon \R^N \to \R^N$~\cite{rfb15} which is employed as a residual network to post-process sparse-view FBP images.
The learnable parameters $\NNparam$ are trained from the collection of $M=4000$ sinogram-image pairs $\{(\y^i, \x^i) \}_{i=1}^M$ provided by the AAPM challenge.
This is achieved by standard empirical risk minimization, i.e., by (approximately) solving
\begin{equation}\label{eq:erm}
	\min_{\NNparam} \ \tfrac{1}{M}\sum_{i=1}^M \norm[\big]{\x^i - \left[\UNet[\NNparam]\circ \FBP\right] (\y^i)}_2^2 + \mu \cdot \norm{\NNparam}_2^2 \ ,
\end{equation}
where we choose $\mu = 10^{-3}$, and $\FBP \colon \R^m \to \R^N$ is obtained from Step~1.
This minimization problem is tackled by $400$ epochs of mini-batch stochastic gradient descent and the Adam optimizer~\cite{kb14} with initial learning rate $0.0002$ and batch size~$4$.

\begin{remark}
	The post-processing strategy of Step 2 was pioneered by~\citet{kmy17,che+17b} and popularized by~\citet{jmfu17,che+17}, among many others.
	Due to the multi-scale encoder-decoder structure with skip-connections, the UNet-architecture is very efficient in handling image-to-image problems.
	Therefore, solving~\eqref{eq:erm} typically works out-of-the-box without requiring sophisticated initialization or optimization strategies (even in seemingly hopeless situations \cite{hauptmann20}).
	Making use of a more powerful or a more memory-efficient network would be beneficial, e.g., see results for the Tiramisu network in Section~\ref{sec:results_analysis}.
	However, we preferred to keep our workflow as simple as possible and therefore decided to stick to the standard UNet as the main computational building block.
\end{remark}

\subsection*{Step 3: Constructing an Iterative Scheme} \label{subsec:itnet}
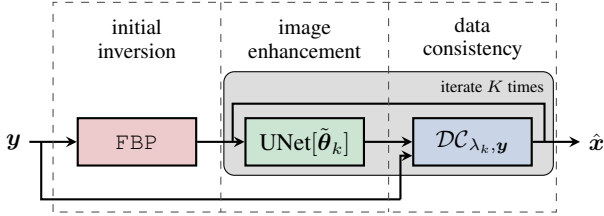
\begin{figure}[t!]
	\centering
	\begin{tikzpicture}[scale=0.9, every node/.style={transform shape}]
    \pgfdeclarelayer{background}
    \pgfsetlayers{background,main}

    \definecolor{deepred}{HTML}{C44E52}
    \definecolor{deepgreen}{HTML}{55A868}
    \definecolor{deepblue}{HTML}{4C72B0}
    \definecolor{deepgray}{HTML}{8C8C8C}

    \node[minimum height=.7cm] (y2) at (0,0) {$\y$};
    \node[thick, draw, fill=deepred!30, minimum width=1.75cm, minimum height=.7cm, right=.7cm of y2] (inv2) {$\FBP$};
    \node[thick, draw, fill=deepgreen!30, minimum width=1.75cm, minimum height=.7cm, right=.7cm of inv2] (net2) {$\UNet[\tilde{\NNparam}_k]$};
    \node[thick, draw, fill=deepblue!30, minimum width=1.75cm, minimum height=.7cm, right=.7cm of net2] (dc2) {$\DC_{\dcparam_k,\y}$};

    \coordinate[above right=.2cm and .175cm of dc2] (abovedc);
    \coordinate[above left=.2cm and .175cm of net2] (abovenet);

    \draw[thick, ->, >=stealth] (y2) -- coordinate[midway](y2inv) coordinate[pos=0.25](y2inv2) (inv2);
    \draw[thick, ->, >=stealth] (inv2) -- coordinate[midway](inv2net) (net2);
    \draw[thick, ->, >=stealth] (net2) -- coordinate[midway](net2dc) coordinate[pos=0.75](net2dc2) (dc2);
    \draw[thick, ->, >=stealth] (net2dc2)+(0,-.85) |- (dc2.193);
    \draw[thick] (net2dc2)+(0,-.85) -| (y2inv2);
    \draw[thick] (abovedc|-dc2) -- (abovedc) -- (abovenet) -- (abovenet|-net2);
    \draw[thick, ->, >=stealth] (dc2.east) -- coordinate[midway] (dc2x) ++ (.7, 0) node[right] {$\xsolu$};

    \node[inner sep=0cm, above=.25cm of abovedc, anchor=east] (it) {\scriptsize iterate $K$ times};
    \node[text height=.5cm, text depth=.3cm, minimum height=1cm, above=.65cm of inv2] (invtext) {\parbox{2cm}{\centering\small initial\\inversion}};
    \node[text height=.5cm, text depth=.3cm, minimum height=1cm, above=.65cm of net2] (nettext) {\parbox{2cm}{\centering\small image\\enhancement}};
    \node[text height=.5cm, text depth=.3cm, minimum height=1cm, above=.65cm of dc2] (dctext) {\parbox{2cm}{\centering\small data\\consistency}};

    \draw[thin, dashed, draw=black!70] (y2inv|-invtext.north) ++ (0, -3.1) rectangle (dc2x|-invtext.north);
    \draw[thin, dashed, draw=black!70] (inv2net|-invtext.north) -- ++ (0, -3.1);
    \draw[thin, dashed, draw=black!70] (net2dc|-invtext.north) -- ++ (0, -3.1);

    \begin{pgfonlayer}{background}
    \node[thin, draw, rounded corners, fill=deepgray!30, fit=(net2) (dc2) (abovedc) (abovenet) (it)] {};
    \end{pgfonlayer}

\end{tikzpicture}
	\vsp{-.1}
	\caption{\unboldmath\textbf{Constructing an iterative scheme.} Schematic reconstruction pipeline of $\ItNet_K[\NNparam]$ defined in \eqref{eq:itnet}.}
	\label{fig:methods:pipeline}
\end{figure}

Our main reconstruction method is called $\ItNet$ (short for \emph{iterative network}).
It incorporates the estimated forward model~$\AFan$ from Step~1 (and the associated inversion $\FBP$) via the following iterative procedure:
\begin{equation}\label{eq:itnet}
	\begin{gathered}
		\ItNet_K[\NNparam] \colon \R^m \to \R^N, \\[1ex]
		\y \mapsto \left[ \bigcirc_{k=1}^K\left( \DC_{\dcparam_k,\y} \circ \UNet[\tilde{\NNparam}_k] \right) \circ \FBP \right](\y),
	\end{gathered}
\end{equation}
for the learnable parameters $\NNparam = \{\tilde{\NNparam}_k,\lambda_k\}_{k=1}^K$, $K\in\N$ and the $k$-th \emph{data-consistency} layer
\begin{equation}
	\DC_{\dcparam_k,\y} \colon \R^N \to \R^N, \ \x \mapsto \x - \dcparam_k \cdot \FBP (\AFan \x - \y).
\end{equation}
The $\ItNet$-architecture\footnote{We drop the subscript $K$ in $\ItNet_K$ whenever it is irrelevant.} is illustrated in Fig.~\ref{fig:methods:pipeline}.
We train it by empirical risk minimization analogously to~\eqref{eq:erm} with $\mu=\nobreak 10^{-4}$.
The UNet-parameters $\tilde{\NNparam}_k$ are initialized by the weights obtained in Step~2.
More details on our precise training approach during the challenge submission phase can be found in Appendix~\ref{sec:exact_challenge_setup}.

We close this section by pointing out several important design choices in our $\ItNet$-architecture:
\begin{romanlist}
	\item
	The centerpiece of $\ItNet$ is the \emph{UNet-architecture}.
	This stands in contrast to earlier generations of unrolled iterative schemes, which rely on basic convolutional blocks instead, e.g., see~\citet{ao18,yslx16}.
	We have found that it is advantageous to exploit the efficacy of UNet-like image-to-image networks as enhancement blocks.
	This is in line with recent state-of-the-art models, which also make use of various advanced sub-networks, e.g., see~\citet{kno+20b,muc+20,hsqdsr19,ramzi20,sriram20}.
	\item
	The initialization of the UNet-parameters $\tilde{\NNparam}_k$ with a \emph{pre-trained model} from Step~2 has led to significant performance gains, regarding both training speed and reconstruction accuracy.
	We refer to Section~\ref{sec:results_analysis} and especially Fig.~\ref{fig:pretrain} for a more details.
	\item
	Our \emph{data-consistency} layer is inspired by a gradient step on the loss $\x \mapsto \tfrac{\dcparam_k}{2} \norm{\AFan \x - \y}_2^2$, which would result in the update $\x \mapsto \x - \dcparam_k \cdot \AFan^\T (\AFan \x - \y)$.
	We depart from this scheme by replacing the unfiltered backprojection~$\AFan^\T$ by its filtered counterpart $\FBP$; cf.~\citet{ding+20,tg21}.
	This modification leads to significantly improved results for two reasons: (a) it counteracts the fact that the unfiltered backprojection is smoothing, and (b) it produces images with pixel values at the right scale.
	Therefore, we interpret the $\ItNet$ as an industry-like iterative CT-algorithm (e.g., see~\citet{willemink19}), rather than a neurally-augmented convex optimization scheme.
\end{romanlist}

\section{Results and Analysis} \label{sec:results_analysis}

This section presents the main findings of our case study.
We begin with several challenge-related experiments, followed by a more in-depth analysis of our method. 

\subsection*{Winning the AAPM Challenge and Beyond}

\begin{table*}[p]
	\caption{\unboldmath\textbf{Average RMSE scores for further evaluation.} ``Challenge FBP'' corresponds to the FBP reconstructions included in the challenge dataset. The method ``$\UNet \circ \FBP$'' corresponds to a post-processing UNet as obtained from Step~2 of Section~\ref{sec:methodology}. For more details on our winning-method ``\textnormal{$\ItNett$ ens.}'' (and its pre-steps ``$\ItNet_4$'' and ``$\ItNett$''), see Appendix~\ref{sec:exact_challenge_setup}.}
	\label{tab:res}
	\centering\footnotesize
	\vsp{.15}
	\begin{tabular}{ccccccccc}
		\toprule
		& \multicolumn{2}{c}{Baselines} & \multicolumn{4}{c}{Our Network Variants} & \multicolumn{2}{c}{Comparison Networks}\\
		\cmidrule(lr){2-3} \cmidrule(lr){4-7} \cmidrule(lr){8-9}
		& Challenge FBP & $\FBP$ & $\UNet \circ \FBP$ & $\ItNet_4$ & $\ItNett$ & $\ItNett$ ens. & $\Tira$ & LPD \\
		\midrule
		RMSE & 5.72e-3 & 3.40e-3 & 3.50e-4 & 1.64e-5 & 1.05e-5 & \textbf{6.42e-6} & 2.24e-4 & 1.24e-4 \\
		\bottomrule
	\end{tabular}
\end{table*}

\begin{figure*}[p]
	\centering
	\begin{tabular}{cc}
		\includegraphics[width=0.45\textwidth]{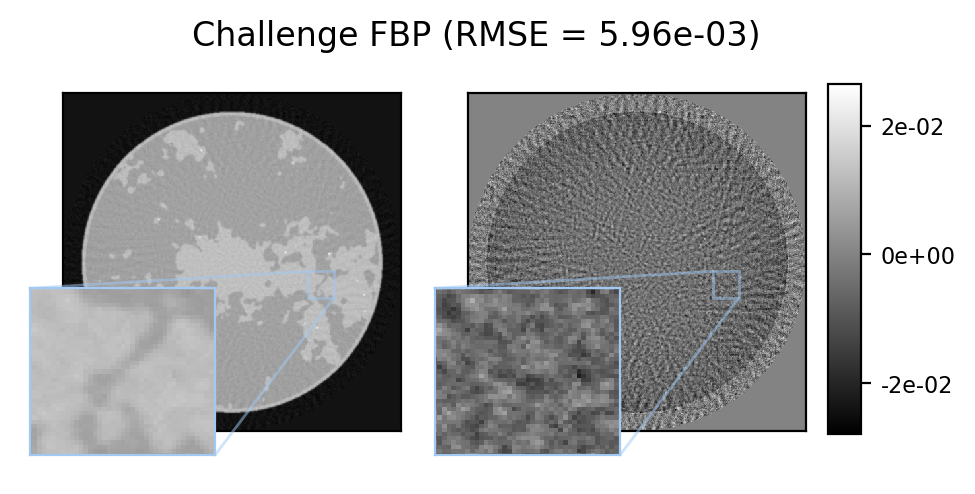} &
		\includegraphics[width=0.45\textwidth]{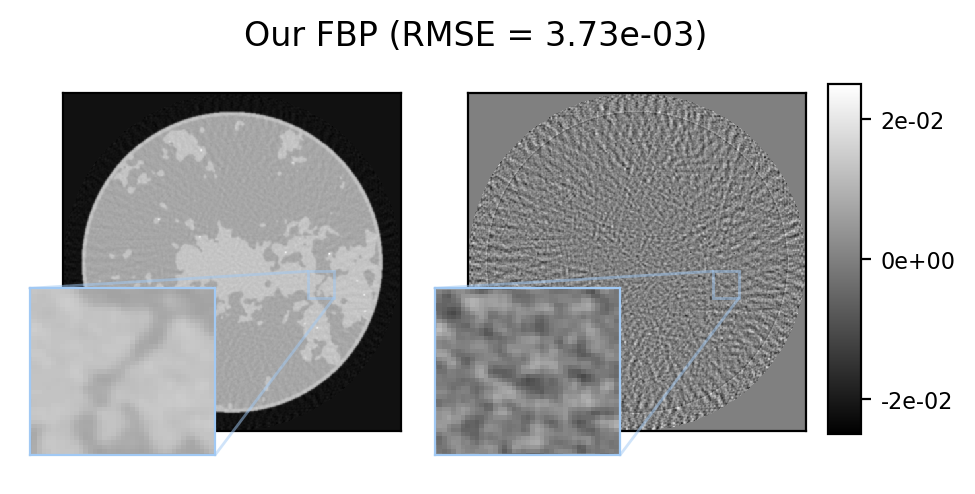} \\
		\includegraphics[width=0.45\textwidth]{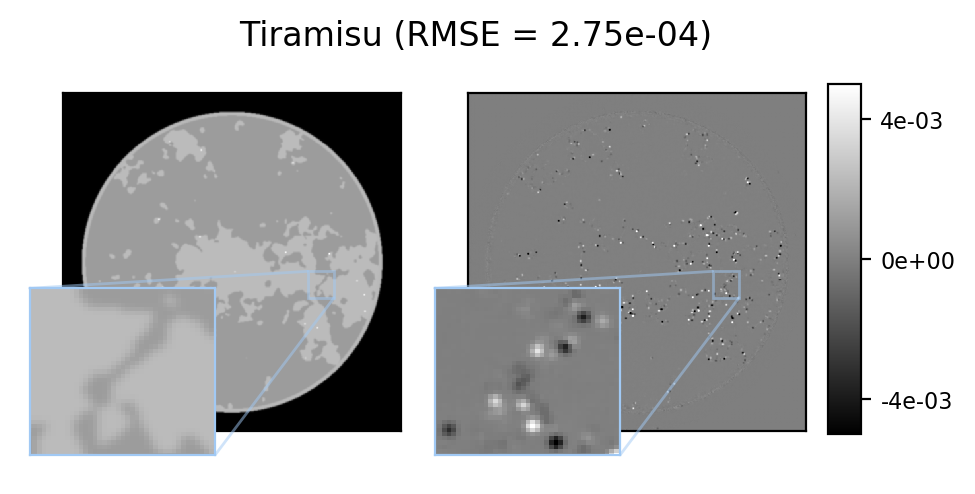} &
		\includegraphics[width=0.45\textwidth]{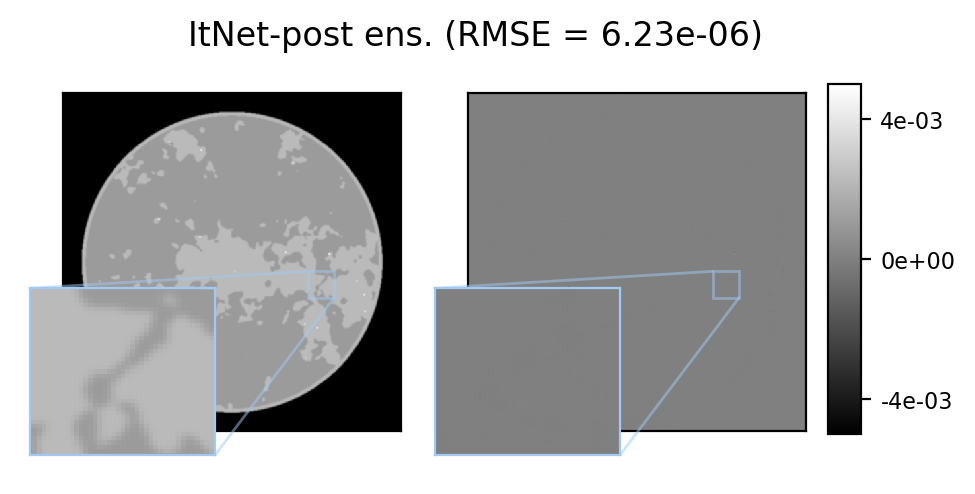}
	\end{tabular}
	\vsp{-.1}
	\caption{\unboldmath\textbf{Reconstruction results.} We display individual reconstructions for an image from the validation set. The first row compares the FBP provided by the AAPM challenge with our FBP ($=\FBP$, see Step~1 of Section~\ref{sec:methodology}). The second row compares a post-processing $\Tira$ with the (ensemble) $\ItNett$. The ground-truth image is omitted because it is visually indistinguishable from the reconstruction of $\ItNett$.}
	\label{fig:recs}
\end{figure*}

\begin{figure*}[p]
	\centering
	\includegraphics[width=\textwidth]{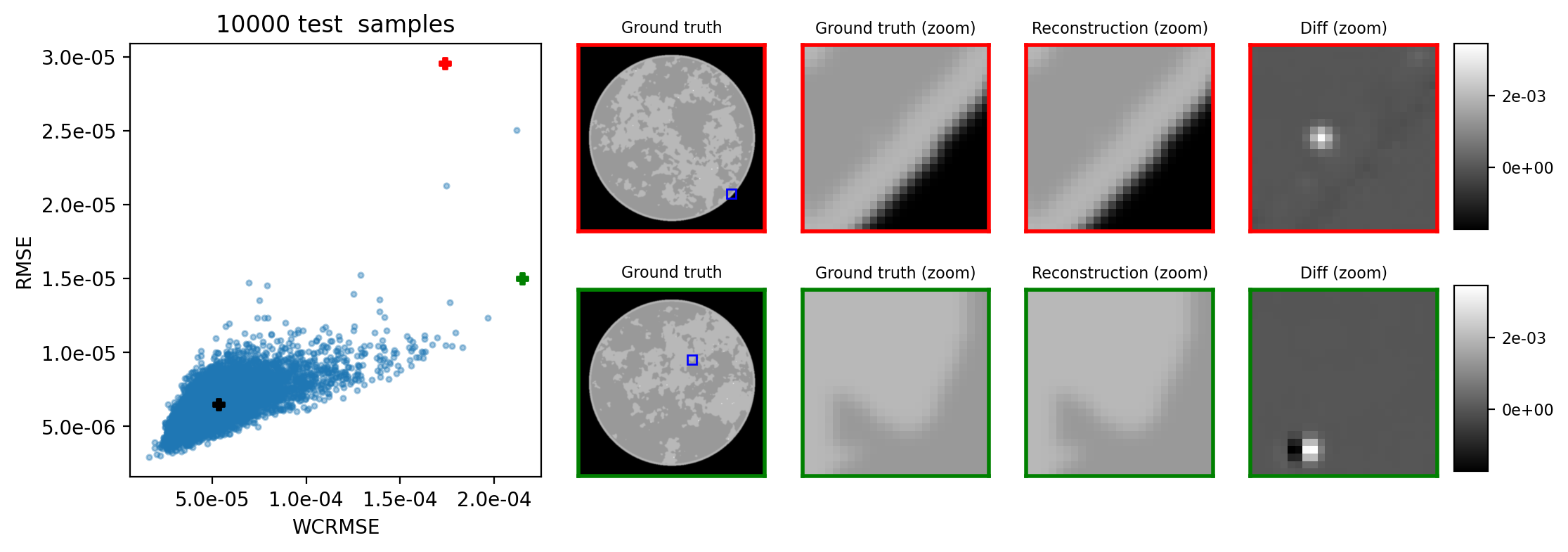}
	\vsp{-.1}
	\caption{\unboldmath\textbf{Consistently accurate?} The plot on the left-hand side visualizes the reconstruction errors of the (ensemble) $\ItNett$ with respect to the RMSE and the WCRMSE (\emph{worst-case} RMSE) over a set of 10000 test images. Note that the WCRMSE was used as secondary challenge metric, computing the highest RMSE value over all $25\times25$ sub-patches of each image. The error distribution indicates a low variance in the reconstruction performance of our method.
		On the right-hand side, we show the worst-case $25\times25$ sub-patches of the images corresponding to the red point (worst RMSE) and green point (worst WCRMSE). The black point represents the average RMSE and WCRMSE.}
	\label{fig:test_data_more}
\end{figure*}

In terms of quantitative similarity measures, we restrict ourselves to reporting the average RMSE, which was the main evaluation metric for the AAPM challenge \cite{sid+21,sp21}.
With an ensembling of ten $\ItNet_5$ (more precisely a variant thereof referred to as $\ItNett$, see Appendix~\ref{sec:exact_challenge_setup}), we were able to achieve near-exact recovery on the test set, thereby winning the challenge with a margin of about an order of magnitude ahead of the runner-up team. The RMSE scores of all participating teams were spread across more than two orders of magnitude in a range between 6.37e-6 (ours) and 7.90e-4.
Remarkably, four out of the five top-performing teams have estimated the forward fanbeam operator and made use of the sinogram data. Two of them computed an approximate TV minimization solution that was further processed by a trained neural network.
The resulting solution maps involve much higher computational costs than our $\ItNett$, due to a significantly larger number of forward-model evaluations.
Note that reaching the first place amounts to a direct comparison with 24 competing methods, see the official AAPM challenge report \cite{sp21} for more details.

Nevertheless, for further analysis, we have benchmarked variants of the $\ItNet$ with different in-house baselines and other state-of-the-art methods.
More specifically, we consider a post-processing of the $\FBP$ by the more powerful \emph{Tiramisu-architecture}~\cite{jdvrb17,bub+19,genzel20} (in comparison to the UNet) as well as the iterative \emph{learned primal-dual} (LPD) algorithm~\cite{ao18} (modified by replacing the unfiltered backprojection with the $\FBP$).
LPD has been recently reported as state-of-the-art in the literature, e.g., see~\citet{ramzi20,leuschner21}.
Table~\ref{tab:res} shows the average RMSE scores for all methods\footnote{Note that we report the RMSE on a subset of 125 images from the training set used for validation. Hence, values differ slightly from the actual results on the official test set. In the final challenge evaluation, $\ItNett$ has achieved an RMSE of \mbox{6.37e-6}.} and Fig.~\ref{fig:recs} visualizes reconstructions of an image from the validation set.

After the competition period, the challenge organizer has provided us with 10000 additional test samples to increase the statistical significance of our evaluation.
The resulting error distribution is visualized in Fig.~\ref{fig:test_data_more}.
Although there are very few outliers, even these reconstructions are visually indistinguishable from the corresponding ground-truth phantoms.
This underscores that the $\ItNett$ solves the CT inverse problem on the given data distribution satisfactorily.

\subsection*{Data-Consistency}
\begin{figure}[t!]
	\centering
	\includegraphics[width=.95\columnwidth]{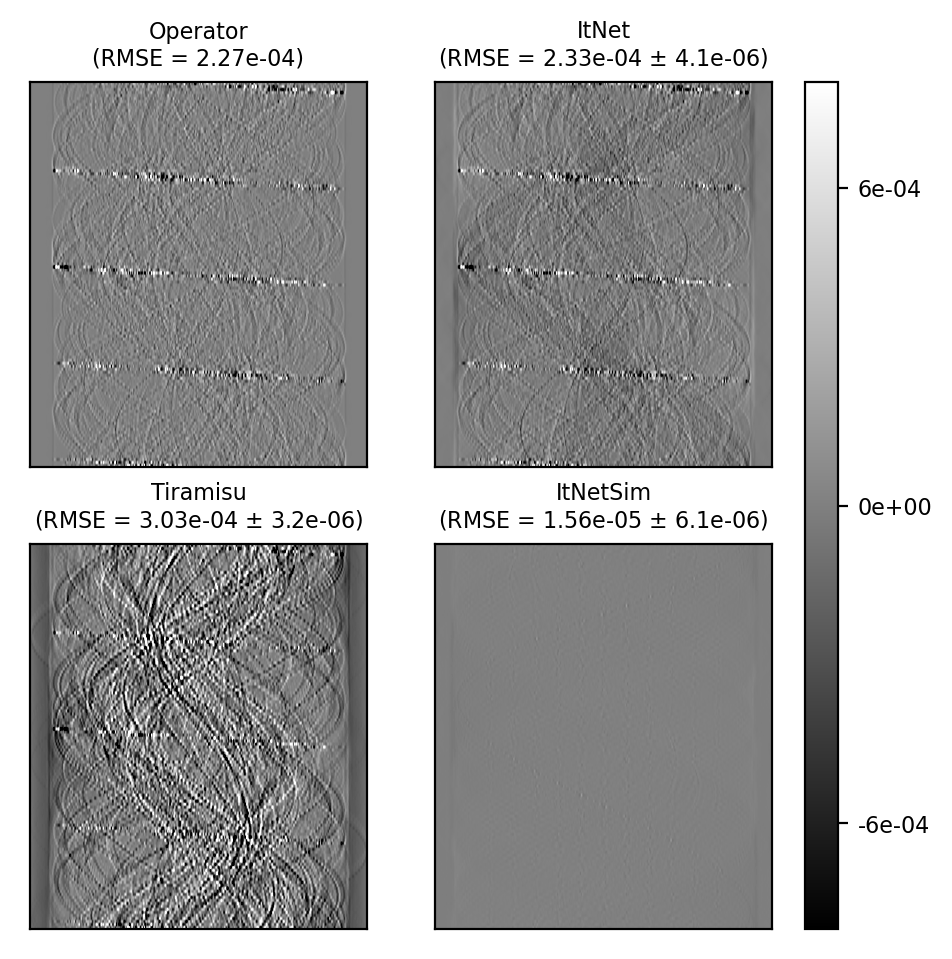}
	\vsp{-.1}
	\caption{\unboldmath\textbf{Data consistency.} We analyze the accuracy of our estimated forward operator by displaying the difference $\y - \AFan \x$ for a sinogram-image pair $(\y,\x)$ from the validation set (top left); the corresponding error is the RMSE averaged over all differences. The difference $\y -\nobreak \AFan \cdot\nobreak \ItNet(\y)$ is visually nearly indistinguishable (top right), showing that $\ItNet$ inherits the inaccuracies from the forward model. Indeed, ItNetSim exhibits a much smaller data-consistency error due to a perfectly matching forward model (bottom right). In contrast, post-processing via Tiramisu (cf.~Table~\ref{tab:res}) reveals a clear lack of data-consistency (bottom left). All images are shown within the same dynamical range.}
	\label{fig:data_consistency}
\end{figure}

A crucial feature of a proper solver for an inverse problem is its consistency with the forward model.
In our case, this means that the difference $\y -\nobreak \AFan \cdot\nobreak \ItNet(\y)$ should be as small as possible.
We analyze this aspect in Fig.~\ref{fig:data_consistency}.\footnote{Here, we have considered an ensemble of five $\ItNet_4$.}
We observe that the data-consistency error is dominated by the error caused by the estimated forward model $\AFan$ (according to Step~1 of Section~\ref{sec:methodology}).
This indicates that the performance of the $\ItNet$ could be further improved if the exact forward operator would be available.
To test this hypothesis, we have trained an $\ItNet$ on sinogram data that was simulated from the ground-truth phantoms using our own estimated forward operator.
As expected, the resulting ItNetSim is more accurate (about factor 2), and according to Fig.~\ref{fig:data_consistency} bottom right, implies a much smaller data-consistency error.
It is also noteworthy that the loss of data-consistency for the Tiramisu is about a factor~20 larger compared to the $\ItNet$,\footnote{This refers to the ratio $\frac{\text{RMSE}(\y,\AFan \cdot \Tira(\y)) - \text{RMSE}(\y,\AFan \x)}{\text{RMSE}(\y,\AFan \cdot \ItNet(\y)) - \text{RMSE}(\y,\AFan \x)}$.} which highlights a typical downside of simple post-processing approaches.

\subsection*{Forward Operator Needed? ... Yes! But How Often?}
\begin{figure*}[t!]
	\centering
	\includegraphics[width=\textwidth]{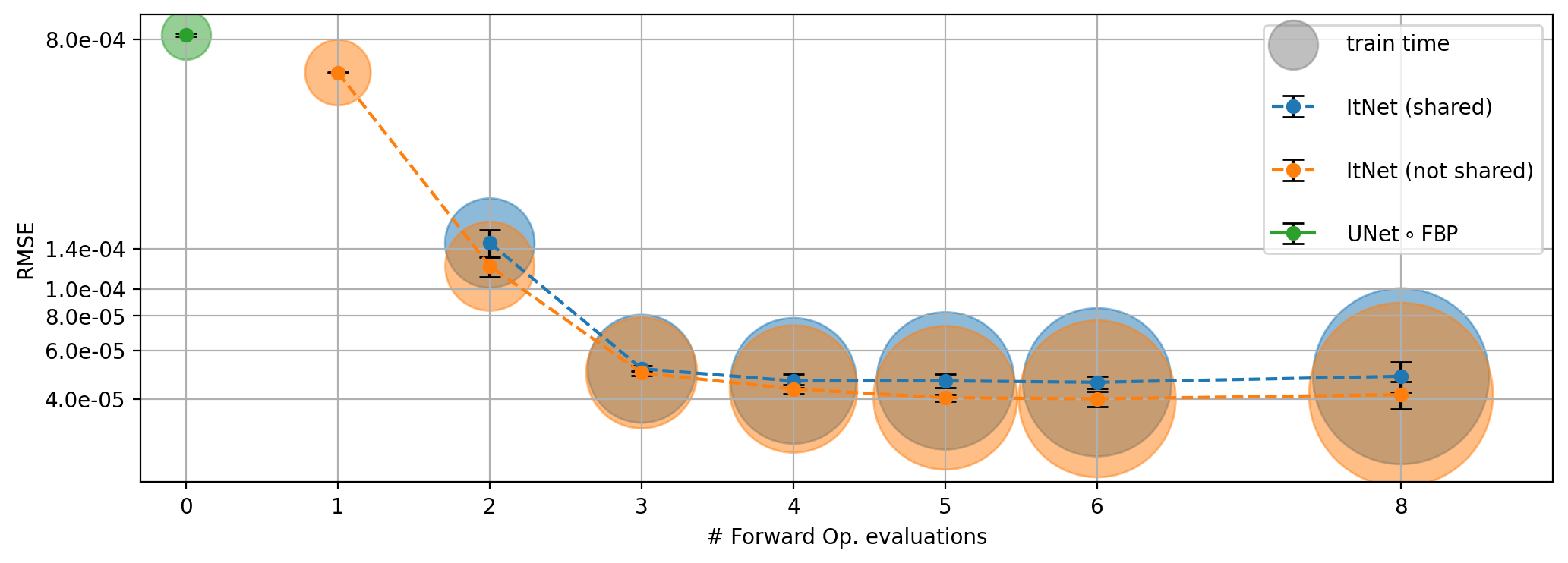}
	\vsp{-.1}
	\caption{\unboldmath\textbf{The deeper the better?} Accuracy of $\ItNet_K$ for different $K$ with (blue) and without (orange) UNet weight sharing. The radii of the circles are proportional to the training time. The mean RMSE ($\pm$ std.~dev.) on a hold-out evaluation data set is reported over 5 different training/validation splits. The original AAPM challenge data was subsampled to the resolution $256 \times 256$ for this experiment.}
	\label{fig:test_over_niter}
\end{figure*}

The previous considerations have particularly demonstrated that incorporating the forward model is key to highly accurate and data-consistent reconstructions.
However, invoking the forward operator often forms the computational bottleneck of a given solution method.
It is therefore important to analyze the effective number of forward/adjoint operator calls required for satisfactory precision.
We address this by training $\ItNet_K$ for different numbers of iterations.\footnote{Due the significant computational effort required to conduct such an experiment, this was done on subsampled $256\times 256$ phantom images and simulated $64$-view sinograms.}
In a nutshell, Fig.~\ref{fig:test_over_niter} confirms that only a few forward operator calls are sufficient for near-exact recovery by the $\ItNet$, which is a notable difference to classical model-based methods like TV-minimization.
A closer look reveals that (a) not sharing the UNet-weights consistently outperforms weight sharing\footnote{This means that the UNet-parameters are shared between all iterations, i.e., enforcing $\tilde \NNparam_1=\dots=\tilde\NNparam_K$ at the training stage.} by a small margin independent of the number of iterations, and (b) there is a sweetspot at about $K=5$ after which the performance gain due to increasing $K$ is negligible and only the training time increases.
For a further discussion of weight sharing, see Appendix~\ref{sec:weightshare}.

\subsection*{Pre-Training Matters}

\begin{figure}[t!]
	\centering
	\includegraphics[width=.75\columnwidth]{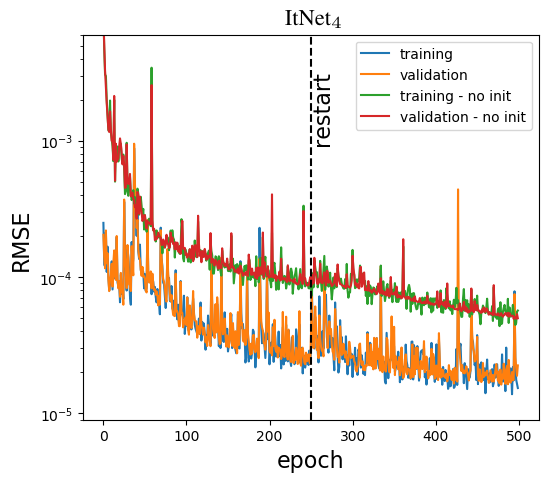}
	\vsp{-.1}
	\caption{\unboldmath\textbf{The power of pre-training.} Loss curves when training the $\ItNet$ with and without a pre-initialization from Step~2 of Section~\ref{sec:methodology}. Note that the above loss curve only corresponds to a part of our full training pipeline, see Fig.~\ref{fig:train} in Appendix~\ref{sec:exact_challenge_setup} for the complete picture.}
	\label{fig:pretrain}
	\vsp{-.1}
\end{figure}

When constructing the $\ItNet$ according to Step~3 of Section~\ref{sec:methodology}, we have observed that it is crucial to initialize the UNet-parameters $\tilde{\NNparam}_k$ by the weights from the post-processing network in Step~2.
This does not only increase the speed of convergence of training $\ItNet$, but it also significantly improves the final accuracy, see Fig.~\ref{fig:pretrain} for corresponding loss curves.
Thus, our results show that the pre-initialization of the UNet-blocks allows finding better local minima.
While the benefits of pre-trained modules are well-known for many standard machine learning tasks, to the best of our knowledge, this has not been reported in the context of inverse problems yet.

\section{Conclusion} \label{sec:discussion}

We have demonstrated that deep-learning-based solvers can produce near-perfect reconstructions for a noise-free CT inverse problem.
While our approach provides evidence of feasibility, several aspects are beyond the scope of this article, some of which are pointed out in the following.

\textbf{How accurate can/should we become?}
The reconstruction error of $\ItNett$ reported in Table~\ref{tab:res} is not exactly zero, yet comparable to the precision of TV minimization, cf.~\citet{sidky20}.
There is no evidence why even more accurate results should not be achievable, for example, by increasing the internal machine precision of \texttt{PyTorch} (which is \mbox{$\approx$1.19e-7} for float32); but this tweak would certainly also affect model-based algorithms.
However, non-perfect recovery is not a severe issue from an applied perspective, since it is typically not required for \emph{practical solutions} to inverse problems.
We believe that our submission to the AAPM challenge has obtained satisfactory results in that respect (i.e., reconstructions visually indistinguishable from the ground-truth phantoms; see Fig.~\ref{fig:test_data_more}).
Having said this, the term ``near-perfect accuracy'' should be used with some care when it comes to real-world scenarios. For instance, realistic CT systems involve analog-to-digital conversion processes and measurement noise, which inevitably leads to reconstruction errors. Therefore, the operating regime of the present paper is primarily a testing ground for exploring the potential capabilities of learning-based methods.

\textbf{Fully data-driven or hybrid method?}
Although the $\ItNet$-architecture is inspired by unrolling, it is not clear to us how well its internal mechanisms match those of classical iterative algorithms.
Indeed, a distinctive feature of our approach is that only very few (five) iterations can achieve near-perfect recovery.
This stands in stark contrast to model-based counterparts, which typically require hundreds or thousands of iterations to converge (resulting in significantly increased computation times).
Therefore, we prefer viewing the $\ItNet$ as fully data-driven pipeline that is \emph{model-aided} by data-consistency terms, rather than a hybrid method; see also \citet{swed20}.
More generally, we suspect that viewing unrolled networks as neurally-enhanced iterative schemes only partially explains the success of deep learning in inverse problems.

\textbf{Model distortions?}
Since the purpose of data-driven methods is to adapt to a specific data distribution, the generalization to out-of-distribution features and forward-model distortions cannot be taken for granted.
This aspect forms a field of active research, e.g., see \citet{arpah20,dch21,gow21b} for initial results, but clearly goes beyond the scope of this paper.

\textbf{Generalization to other (inverse) problems?}
The AAPM challenge has provided an ideal experimental area to test our research hypothesis.
Although this has enabled an insightful reliability check, a foundational understanding of learned reconstruction methods is still in its infancy.
In particular, it remains speculative to what extent our findings would generalize beyond the sparse-view, mono-energy CT setup.
Therefore, similar case studies for different types of inverse problems and real-world data are important steps for future research.
Our evaluation on the more realistic LoDoPaB CT dataset in Appendix~\ref{sec:lodopab} can be seen as a first effort in that direction.

\section*{Acknowledgements}

We would like to thank Emil Sidky for very fruitful discussions and his encouraging support.
Moreover, we express our gratitude to the organizers of the AAPM challenge: Emil Sidky, Xiaochuan Pan, Jovan Brankov, Iris Lorente, Samuel Armato, and the AAPM Working Group on Grand Challenges. I.~G.~would like to thank Klaus-Robert~M{\"u}ller for valuable suggestions to improve the manuscript.

\bibliography{paper-arxiv}

\begin{thebibliography}{50}
\providecommand{\natexlab}[1]{#1}
\providecommand{\url}[1]{\texttt{#1}}
\expandafter\ifx\csname urlstyle\endcsname\relax
  \providecommand{\doi}[1]{doi: #1}\else
  \providecommand{\doi}{doi: \begingroup \urlstyle{rm}\Url}\fi

\bibitem[Adler \& {\"O}ktem(2018)Adler and {\"O}ktem]{ao18}
Adler, J. and {\"O}ktem, O.
\newblock Learned primal-dual reconstruction.
\newblock \emph{IEEE Trans. Med. Imag.}, 37\penalty0 (6):\penalty0 1322--1332,
  2018.

\bibitem[Aggarwal et~al.(2018)Aggarwal, Mani, and Jacob]{amj18}
Aggarwal, H.~K., Mani, M.~P., and Jacob, M.
\newblock {MoDL:} model-based deep learning architecture for inverse problems.
\newblock \emph{IEEE Trans. Med. Imag.}, 38\penalty0 (2):\penalty0 394--405,
  2018.

\bibitem[Antun et~al.(2020)Antun, Renna, Poon, Adcock, and Hansen]{arpah20}
Antun, V., Renna, F., Poon, C., Adcock, B., and Hansen, A.~C.
\newblock On instabilities of deep learning in image reconstruction and the
  potential costs of {AI}.
\newblock \emph{Proc. Natl. Acad. Sci.}, 117\penalty0 (48):\penalty0
  30088--30095, 2020.

\bibitem[Arridge et~al.(2019)Arridge, Maass, {\"O}ktem, and
  Sch{\"o}nlieb]{amos19}
Arridge, S., Maass, P., {\"O}ktem, O., and Sch{\"o}nlieb, C.-B.
\newblock Solving inverse problems using data-driven models.
\newblock \emph{Acta Numer.}, 28:\penalty0 1--174, 2019.

\bibitem[Bubba et~al.(2019)Bubba, Kutyniok, Lassas, M{\"a}rz, Samek, Siltanen,
  and Srinivasan]{bub+19}
Bubba, T.~A., Kutyniok, G., Lassas, M., M{\"a}rz, M., Samek, W., Siltanen, S.,
  and Srinivasan, V.
\newblock Learning the invisible: A hybrid deep learning-shearlet framework for
  limited angle computed tomography.
\newblock \emph{Inverse Probl.}, 35\penalty0 (6):\penalty0 064002, 2019.

\bibitem[Cand{\`e}s et~al.(2006)Cand{\`e}s, Romberg, and Tao]{crt06a}
Cand{\`e}s, E.~J., Romberg, J.~K., and Tao, T.
\newblock Robust uncertainty principles: exact signal reconstruction from
  highly incomplete frequency information.
\newblock \emph{IEEE Trans. Inf. Theory}, 52\penalty0 (2):\penalty0 489--509,
  2006.

\bibitem[Chen et~al.(2017{\natexlab{a}})Chen, Zhang, Kalra, Lin, Chen, Liao,
  Zhou, and Wang]{che+17}
Chen, H., Zhang, Y., Kalra, M.~K., Lin, F., Chen, Y., Liao, P., Zhou, J., and
  Wang, G.
\newblock Low-dose {CT} with a residual encoder-decoder convolutional neural
  network.
\newblock \emph{IEEE Trans. Med. Imag.}, 36\penalty0 (12):\penalty0 2524--2535,
  2017{\natexlab{a}}.

\bibitem[Chen et~al.(2017{\natexlab{b}})Chen, Zhang, Zhang, Liao, Li, Zhou, and
  Wang]{che+17b}
Chen, H., Zhang, Y., Zhang, W., Liao, P., Li, K., Zhou, J., and Wang, G.
\newblock Low-dose {CT} via convolutional neural network.
\newblock \emph{Biomed. Opt. Express}, 8\penalty0 (2):\penalty0 679--694,
  2017{\natexlab{b}}.

\bibitem[Chen et~al.(2018)Chen, Zhang, Chen, Zhang, Zhang, Sun, Lv, Liao, Zhou,
  and Wang]{che+18}
Chen, H., Zhang, Y., Chen, Y., Zhang, J., Zhang, W., Sun, H., Lv, Y., Liao, P.,
  Zhou, J., and Wang, G.
\newblock {LEARN:} learned experts' assessment-based reconstruction network for
  sparse-data ct.
\newblock \emph{IEEE Trans. Med. Imag.}, 37\penalty0 (6):\penalty0 1333--1347,
  2018.

\bibitem[{Chun} et~al.(2020){Chun}, {Huang}, {Lim}, and {Fessler}]{chlf+20}
{Chun}, I.~Y., {Huang}, Z., {Lim}, H., and {Fessler}, J.
\newblock {Momentum-Net:} fast and convergent iterative neural network for
  inverse problems.
\newblock \emph{IEEE Trans. Pattern Anal. Mach. Intell.}, available online:
  \url{https://doi.org/10.1109/TPAMI.2020.3012955}, 2020.

\bibitem[Darestani et~al.(2021)Darestani, Chaudhari, and Heckel]{dch21}
Darestani, M.~Z., Chaudhari, A., and Heckel, R.
\newblock Measuring robustness in deep learning based compressive sensing.
\newblock In \emph{Proceedings of the 38th International Conference on Machine
  Learning (ICML)}, pp.\  2433--2444, 2021.

\bibitem[Ding et~al.(2020)Ding, Chen, Zhang, Huang, Ji, and Gao]{ding+20}
Ding, Q., Chen, G., Zhang, X., Huang, Q., Ji, H., and Gao, H.
\newblock Low-dose {CT} with deep learning regularization via proximal
  forward{\textendash}backward splitting.
\newblock \emph{Phys. Med. Biol.}, 65\penalty0 (12):\penalty0 125009, 2020.

\bibitem[Donoho(2006)]{don06}
Donoho, D.~L.
\newblock Compressed sensing.
\newblock \emph{IEEE Trans. Inf. Theory}, 52\penalty0 (4):\penalty0 1289--1306,
  2006.

\bibitem[Fessler(2017)]{fessler17}
Fessler, J.~A.
\newblock Analytical tomographic image reconstruction methods (chapter 3 of
  book draft).
\newblock URL: \url{https://web.eecs.umich.edu/~fessler/book/c-tomo.pdf}, 2017.

\bibitem[Foucart \& Rauhut(2013)Foucart and Rauhut]{fh13}
Foucart, S. and Rauhut, H.
\newblock \emph{A Mathematical Introduction to Compressive Sensing}.
\newblock Applied and Numerical Harmonic Analysis. Birkh{\"a}user Basel, 2013.

\bibitem[Genzel et~al.(2022)Genzel, Macdonald, and M{\"a}rz]{genzel20}
Genzel, M., Macdonald, J., and M{\"a}rz, M.
\newblock {Solving Inverse Problems With Deep Neural Networks -- Robustness
  Included?}
\newblock \emph{IEEE Trans. Pattern Anal. Mach. Intell.}, available online:
  \url{https://doi.org/10.1109/tpami.2022.3148324}, 2022.

\bibitem[Gilton et~al.(2021{\natexlab{a}})Gilton, Ongie, and Willett]{gow21}
Gilton, D., Ongie, G., and Willett, R.
\newblock Deep equilibrium architectures for inverse problems in imaging.
\newblock \emph{IEEE Trans. Comput. Imag.}, 7:\penalty0 1123--1133,
  2021{\natexlab{a}}.

\bibitem[Gilton et~al.(2021{\natexlab{b}})Gilton, Ongie, and Willett]{gow21b}
Gilton, D., Ongie, G., and Willett, R.
\newblock Model adaptation for inverse problems in imaging.
\newblock \emph{IEEE Trans. Comput. Imag.}, 7:\penalty0 661--674,
  2021{\natexlab{b}}.

\bibitem[Goodfellow et~al.(2016)Goodfellow, Bengio, and Courville]{gbc16}
Goodfellow, I., Bengio, Y., and Courville, A.
\newblock \emph{Deep Learning}.
\newblock MIT Press, 2016.

\bibitem[Gregor \& LeCun(2010)Gregor and LeCun]{kl10}
Gregor, K. and LeCun, Y.
\newblock Learning fast approximations of sparse coding.
\newblock In \emph{Proceedings of the 27th International Conference on Machine
  Learning (ICML)}, pp.\  399--406, 2010.

\bibitem[Hammernik et~al.(2018)Hammernik, Klatzer, Kobler, Recht, Sodickson,
  Pock, and Knoll]{ham+17}
Hammernik, K., Klatzer, T., Kobler, E., Recht, M.~P., Sodickson, D.~K., Pock,
  T., and Knoll, F.
\newblock Learning a variational network for reconstruction of accelerated
  {MRI} data.
\newblock \emph{Magn. Reson. Med.}, 79\penalty0 (6):\penalty0 3055--3071, 2018.

\bibitem[Hammernik et~al.(2021)Hammernik, Schlemper, Qin, Duan, Summers, and
  Rueckert]{hsqdsr19}
Hammernik, K., Schlemper, J., Qin, C., Duan, J., Summers, R.~M., and Rueckert,
  D.
\newblock Systematic evaluation of iterative deep neural networks for fast
  parallel mri reconstruction with sensitivity-weighted coil combination.
\newblock \emph{Magn. Reson. Med.}, 86\penalty0 (4):\penalty0 1859--1872, 2021.

\bibitem[Hauptmann \& Adler(2020)Hauptmann and Adler]{hauptmann20}
Hauptmann, A. and Adler, J.
\newblock On the unreasonable effectiveness of {CNNs}.
\newblock Preprint arXiv:2007.14745, 2020.

\bibitem[Heaton et~al.(2021)Heaton, Fung, Gibali, and Yin]{hfgy21}
Heaton, H., Fung, S.~W., Gibali, A., and Yin, W.
\newblock Feasibility-based fixed point networks.
\newblock arXiv:2104.14090, 2021.

\bibitem[J{\'e}gou et~al.(2017)J{\'e}gou, Drozdzal, Vazquez, Romero, and
  Bengio]{jdvrb17}
J{\'e}gou, S., Drozdzal, M., Vazquez, D., Romero, A., and Bengio, Y.
\newblock The one hundred layers tiramisu: Fully convolutional densenets for
  semantic segmentation.
\newblock In \emph{Proceedings of the IEEE Conference on Computer Vision and
  Pattern Recognition (CVPR)}, pp.\  11--19, 2017.

\bibitem[Jin et~al.(2017)Jin, McCann, Froustey, and Unser]{jmfu17}
Jin, K.~H., McCann, M.~T., Froustey, E., and Unser, M.
\newblock Deep convolutional neural network for inverse problems in imaging.
\newblock \emph{IEEE Trans. Image Process.}, 26\penalty0 (9):\penalty0
  4509--4522, 2017.

\bibitem[Kang et~al.(2017)Kang, Min, and Ye]{kmy17}
Kang, E., Min, J., and Ye, J.~C.
\newblock A deep convolutional neural network using directional wavelets for
  low-dose {X-ray CT} reconstruction.
\newblock \emph{Med. Phys.}, 44\penalty0 (10):\penalty0 e360--e375, 2017.

\bibitem[Kingma \& Ba(2014)Kingma and Ba]{kb14}
Kingma, D.~P. and Ba, J.
\newblock {Adam:} a method for stochastic optimization.
\newblock Preprint arXiv:1412.6980, 2014.

\bibitem[Knoll et~al.(2020)Knoll, Murrell, Sriram, Yakubova, Zbontar, Rabbat,
  Defazio, Muckley, Sodickson, Zitnick, and Recht]{kno+20b}
Knoll, F., Murrell, T., Sriram, A., Yakubova, N., Zbontar, J., Rabbat, M.,
  Defazio, A., Muckley, M.~J., Sodickson, D.~K., Zitnick, C.~L., and Recht,
  M.~P.
\newblock Advancing machine learning for {MR} image reconstruction with an open
  competition: Overview of the 2019 {fastMRI} challenge.
\newblock \emph{Magn. Reson. Med.}, 84\penalty0 (6):\penalty0 3054--3070, 2020.

\bibitem[LeCun et~al.(2015)LeCun, Bengio, and Hinton]{lbh15}
LeCun, Y., Bengio, Y., and Hinton, G.
\newblock Deep learning.
\newblock \emph{Nature}, 521\penalty0 (7553):\penalty0 436--444, 2015.

\bibitem[Leuschner et~al.(2021)Leuschner, Schmidt, Ganguly, Andriiashen, Coban,
  Denker, Bauer, Hadjifaradji, Batenburg, Maass, and van
  Eijnatten]{leuschner21}
Leuschner, J., Schmidt, M., Ganguly, P.~S., Andriiashen, V., Coban, S.~B.,
  Denker, A., Bauer, D., Hadjifaradji, A., Batenburg, K.~J., Maass, P., and van
  Eijnatten, M.
\newblock Quantitative comparison of deep learning-based image reconstruction
  methods for low-dose and sparse-angle {CT} applications.
\newblock \emph{J. Imaging}, 7\penalty0 (3), 2021.

\bibitem[Liu et~al.(2020)Liu, Jiang, He, Chen, Liu, Gao, and Han]{RAdam}
Liu, L., Jiang, H., He, P., Chen, W., Liu, X., Gao, J., and Han, J.
\newblock On the variance of the adaptive learning rate and beyond.
\newblock In \emph{International Conference on Learning Representations
  (ICLR)}, 2020.

\bibitem[Loshchilov \& Hutter(2019)Loshchilov and Hutter]{AdamW}
Loshchilov, I. and Hutter, F.
\newblock Decoupled weight decay regularization.
\newblock In \emph{International Conference on Learning Representations
  (ICLR)}, 2019.

\bibitem[Muckley et~al.(2020)Muckley, Riemenschneider, Radmanesh, Kim, Jeong,
  Ko, Jun, Shin, Hwang, Mostapha, Arberet, Nickel, Ramzi, Ciuciu, Starck,
  Teuwen, Karkalousos, Zhang, Sriram, Huang, Yakubova, Lui, and Knoll]{muc+20}
Muckley, M.~J., Riemenschneider, B., Radmanesh, A., Kim, S., Jeong, G., Ko, J.,
  Jun, Y., Shin, H., Hwang, D., Mostapha, M., Arberet, S., Nickel, D., Ramzi,
  Z., Ciuciu, P., Starck, J.-L., Teuwen, J., Karkalousos, D., Zhang, C.,
  Sriram, A., Huang, Z., Yakubova, N., Lui, Y., and Knoll, F.
\newblock State-of-the-art machine learning {MRI} reconstruction in 2020:
  Results of the second {fastMRI} challenge.
\newblock Preprint arXiv:2012.06318, 2020.

\bibitem[Ongie et~al.(2020)Ongie, Jalal, Baraniuk, Metzler, Dimakis, and
  Willett]{ojbmdw20}
Ongie, G., Jalal, A., Baraniuk, R.~G., Metzler, C.~A., Dimakis, A.~G., and
  Willett, R.
\newblock Deep learning techniques for inverse problems in imaging.
\newblock \emph{IEEE J. Sel. Areas Inf. Theory}, 1\penalty0 (1):\penalty0
  39--56, 2020.

\bibitem[Putzky \& Welling(2017)Putzky and Welling]{pw17}
Putzky, P. and Welling, M.
\newblock Recurrent inference machines for solving inverse problems.
\newblock Preprint arXiv:1706.04008, 2017.

\bibitem[Ramzi et~al.(2020)Ramzi, Ciuciu, and Starck]{ramzi20}
Ramzi, Z., Ciuciu, P., and Starck, J.-L.
\newblock {XPDNet for MRI reconstruction: an application to the fastMRI 2020
  brain challenge}.
\newblock Preprint arXiv:2010.07290, 2020.

\bibitem[Ronneberger et~al.(2015)Ronneberger, Fischer, and Brox]{rfb15}
Ronneberger, O., Fischer, P., and Brox, T.
\newblock {U-Net:} convolutional networks for biomedical image segmentation.
\newblock In Navab, N., Hornegger, J., Wells, W.~M., and Frangi, A.~F. (eds.),
  \emph{Medical Image Computing and Computer Assisted Intervention -- MICCAI
  2015}, pp.\  234--241. Springer Cham, 2015.

\bibitem[Schlemper et~al.(2019)Schlemper, Oksuz, Clough, Duan, King, Schnabel,
  Hajnal, and Rueckert]{sch+19}
Schlemper, J., Oksuz, I., Clough, J.~R., Duan, J., King, A.~P., Schnabel,
  J.~A., Hajnal, J.~V., and Rueckert, D.
\newblock \mbox{dAUTOMAP:} decomposing \mbox{AUTOMAP} to achieve scalability
  and enhance performance.
\newblock Preprint arXiv:1909.10995, 2019.

\bibitem[Schmidhuber(2015)]{sch15}
Schmidhuber, J.
\newblock Deep learning in neural networks: An overview.
\newblock \emph{Neural Netw.}, 61:\penalty0 85--117, 2015.

\bibitem[Shlezinger et~al.(2020)Shlezinger, Whang, Eldar, and Dimakis]{swed20}
Shlezinger, N., Whang, J., Eldar, Y.~C., and Dimakis, A.~G.
\newblock Model-based deep learning.
\newblock Preprint arXiv:2012.08405, 2020.

\bibitem[Sidky et~al.(2021{\natexlab{a}})Sidky, Lorente, Brankov, and
  Pan]{sidky20}
Sidky, E., Lorente, I., Brankov, J.~G., and Pan, X.
\newblock Do {CNNs} solve the {CT} inverse problem?
\newblock \emph{IEEE Trans. Biomed. Eng.}, 68\penalty0 (6):\penalty0
  1799--1810, 2021{\natexlab{a}}.

\bibitem[Sidky et~al.(2021{\natexlab{b}})Sidky, Pan, Brankov, Lorente, Armato,
  Drukker, Hadjiyski, Petrick, Farahani, Munbodh, Cha, Kalpathy-Cramer, Bearce,
  and {AAPM Working Group on Grand challenges}]{sid+21}
Sidky, E., Pan, X., Brankov, J., Lorente, I., Armato, S., Drukker, K.,
  Hadjiyski, L., Petrick, N., Farahani, K., Munbodh, R., Cha, K.,
  Kalpathy-Cramer, J., Bearce, B., and {AAPM Working Group on Grand
  challenges}.
\newblock {Deep Learning for Inverse Problems: Sparse-View Computed Tomography
  Image Reconstruction (DL-sparse-view CT)}.
\newblock URL: \url{https://www.aapm.org/GrandChallenge/DL-sparse-view-CT/},
  2021{\natexlab{b}}.

\bibitem[Sidky \& Pan(2021)Sidky and Pan]{sp21}
Sidky, E.~Y. and Pan, X.
\newblock {Report on the AAPM deep-learning sparse-view CT (DL-sparse-view CT)
  Grand Challenge}.
\newblock \emph{Med. Phys.}, accepted, preprint arXiv:2109.09640, 2021.

\bibitem[Sriram et~al.(2020)Sriram, Zbontar, Murrell, Defazio, Zitnick,
  Yakubova, Knoll, and Johnson]{sriram20}
Sriram, A., Zbontar, J., Murrell, T., Defazio, A., Zitnick, C.~L., Yakubova,
  N., Knoll, F., and Johnson, P.
\newblock End-to-end variational networks for accelerated {MRI} reconstruction.
\newblock In \emph{International Conference on Medical Image Computing and
  Computer-Assisted Intervention}, pp.\  64--73, 2020.

\bibitem[Tirer \& Giryes(2021)Tirer and Giryes]{tg21}
Tirer, T. and Giryes, R.
\newblock On the convergence rate of projected gradient descent for a
  back-projection based objective.
\newblock \emph{SIAM J. Imag. Sci.}, 14\penalty0 (4):\penalty0 1504--1531,
  2021.

\bibitem[Willemink \& No{\"e}l(2019)Willemink and No{\"e}l]{willemink19}
Willemink, M.~J. and No{\"e}l, P.~B.
\newblock The evolution of image reconstruction for {CT} -- from filtered back
  projection to artificial intelligence.
\newblock \emph{Eur. Radiol.}, 29\penalty0 (5):\penalty0 2185--2195, 2019.

\bibitem[Wu \& He(2018)Wu and He]{wu18}
Wu, Y. and He, K.
\newblock Group normalization.
\newblock In \emph{Proceedings of the European conference on computer vision
  (ECCV)}, pp.\  3--19, 2018.

\bibitem[W{\"u}rfl et~al.(2016)W{\"u}rfl, Ghesu, Christlein, and
  Maier]{wurfl16}
W{\"u}rfl, T., Ghesu, F.~C., Christlein, V., and Maier, A.
\newblock Deep learning computed tomography.
\newblock In \emph{Medical Image Computing and Computer-Assisted Intervention
  -- MICCAI 2016}, pp.\  432--440, 2016.

\bibitem[Yang et~al.(2016)Yang, Sun, Li, and Xu]{yslx16}
Yang, Y., Sun, J., Li, H., and Xu, Z.
\newblock Deep {ADMM-Net} for compressive sensing {MRI}.
\newblock In \emph{Advances in Neural Information Processing Systems 29}, pp.\
  10--18, 2016.

\end{thebibliography}
\bibliographystyle{icml2022}

\newpage
\appendix
\onecolumn
\section{Exact AAPM Challenge Setup}\label{sec:exact_challenge_setup}

To ensure reproducability, we give an exact account of how we trained our winning submission to the AAPM challenge (team-name: \texttt{robust-and-stable}).
Since the systematic investigation of the $\ItNet$-architecture was conducted after the challenge submission phase, it became clear that not all of the substeps outlined below have a notable impact on the performance (see also Section~\ref{sec:results_analysis}).

The following details are related to Step~3 of Section~\ref{sec:methodology} (``Constructing an Iterative Scheme'').
We start by training an $\ItNet_4$ (with weight sharing) for $500$ epochs of mini-batch stochastic gradient descent and Adam with an initial learning rate of $8\cdot 10^{-5}$ and a batch size of $2$ (restarting Adam after $250$ epochs).
Then, we improve the accuracy by the following \emph{post-training} strategy:
First, the $\ItNet_4$ is extended by one more iteration:
\begin{equation}
	\label{eq:itnet-5}
	\begin{gathered}
		\ItNett[\NNparam] \colon \R^m \to \R^N,\\[1ex]
		\y \mapsto \left[ \bigcirc_{k=1}^5  \left( \DC_{\dcparam_k,\y} \circ \UNet[\tilde{\NNparam}_k] \right) \circ \FBP \right](\y),
	\end{gathered}
\end{equation}
where $\tilde{\NNparam}_k$ is initialized with the optimized weights from $\ItNet_4$ for $k=1,\dots,4$, and we set $\tilde{\NNparam}_5 \coloneqq \tilde{\NNparam}_4$.
Next, $\ItNett$ is fine-tuned by keeping the weights $\tilde{\NNparam}_1 = \tilde{\NNparam}_2 = \tilde{\NNparam}_3$ of the first three UNet-blocks fixed and optimizing only over the weights of the last two iterations (without weight sharing).

Aiming at an additional training speed-up, we use the initialization $\DCparam = [1.1, 1.3, 1.4, 0.08]$ for the data-consistency parameters of $\ItNet_4$, which was found by pre-training.
Similarly, $\ItNett$ is initialized with the optimized values from $\ItNet_4$ for $k=1,2,3$, together with $\dcparam_4=1.0$ and $\dcparam_5=0.1$.

\begin{figure*}[t!]
	\centering
	\includegraphics[width=\textwidth]{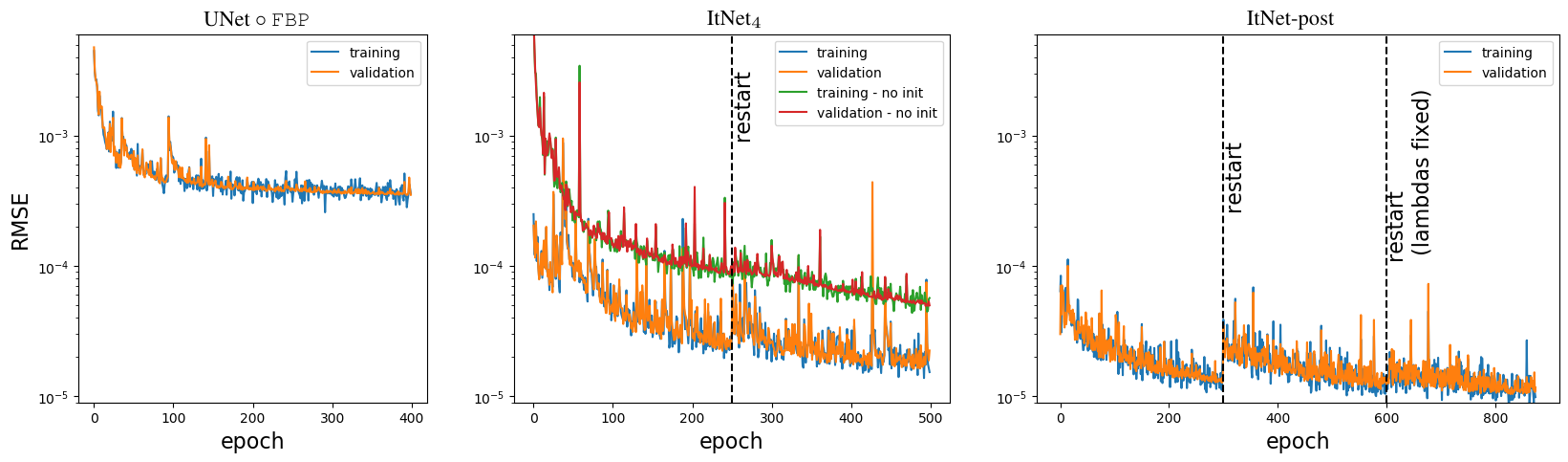}
	\vsp{-.1}
	\caption{\unboldmath\textbf{Loss curves and network training.} The first two plots demonstrate that $\ItNet_4$ improves the RMSE by approximately an order of magnitude in comparison to a post-processing by $\UNet$. Furthermore, the gain of our UNet-initialization strategy can be seen in the second graph. The last two plots illustrate the advantages of restarting and of the post-training strategy, respectively. Note that we display the RMSE on the training and validation sets instead of the actual $\ell^2$-losses, which behave similarly.}
	\label{fig:train}
\end{figure*}

To improve the overall performance of our networks, we have additionally applied the following ``tricks'' for \emph{fine tuning}, which are ordered by their importance:
\begin{romanlist}
	\item
	Due to statistical fluctuations, the networks typically exhibit slightly different reconstruction errors, despite using the same training pipeline.
	The final reconstructions are therefore computed by an \emph{ensemble} of ten networks, each trained on a different split of the training set.
	\item
	Due to the training with small batch sizes, we replace batch normalization of the UNet-architecture by \emph{group normalization}~\cite{wu18}.
	\item
	We equip the UNet-blocks with a few \emph{memory channels}, i.e., one actually has that $\UNet[\NNparam] \colon \R^N \times\nobreak (\R^N)^{c_{\textrm{mem}}} \to \R^N \times\nobreak (\R^N)^{c_{\textrm{mem}}}$; cf.~\citet{pw17,ao18}. While the original image-enhancement channel is not altered, the output of the additional channels is propagated through the $\ItNet$, playing the role of a hidden state (in the spirit of recurrent neural networks). For our experiments, we use $c_{\textrm{mem}} = 5$.
	\item
	It was beneficial to restart occasionally the training of the networks (see also Fig.~\ref{fig:train}).
\end{romanlist}
The following modifications did not lead to a gain in performance and were omitted:
\begin{romanlist}
	\item
	Improving $\FBP$ in Step~1 of Section~\ref{sec:methodology} by making some of it components learnable (e.g., the filter), cf.~\citet{wurfl16}.
	Although this is advantageous for the reconstruction quality of $\FBP$ itself, it leads to worse results for the $\ItNet$.
	This suggests that a combination of model- and data-based methods benefits most from  precise and unaltered physical models.
	\item
	Adding additional convolutional-blocks in the measurement domain of $\ItNet$.
	\item
	Modifying the standard $\ell^2$-loss by incorporating the RMSE or the $\ell^1$-norm.
	\item
	Utilizing different optimizers such as SGD, RAdam~\cite{RAdam}, or AdamW~\cite{AdamW}.
\end{romanlist}
In Fig.~\ref{fig:train}, we visualize the RMSE loss curves of our training pipeline, i.e.,
\begin{equation*}
	\UNet \circ \FBP \to \ItNet_4 (+ \text{restart}) \to \ItNett (+ 2\times\text{restart}).
\end{equation*}

\hfill

\section{The Effect of Weight Sharing}
\label{sec:weightshare}

\begin{figure}[h!]
	\centering
	\includegraphics[width=.5\columnwidth]{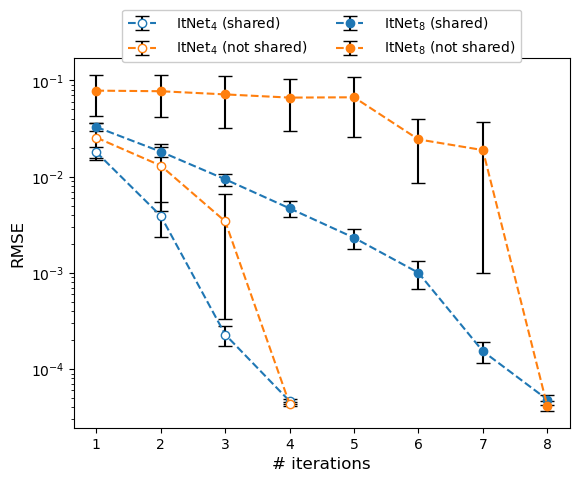}
	\vsp{-.1}
	\caption{\unboldmath\textbf{A look inside.} Accuracy of $\ItNet_4$ and $\ItNet_8$ with (blue) and without (orange) UNet weight sharing when using only the first $k$ iteration steps and discarding the rest. The mean RMSE ($\pm$ std.~dev.) on a hold-out evaluation data set is reported over 5 different training/validation splits. The original AAPM challenge data was subsampled to the resolution $256 \times 256$ for this experiment.}
	\label{fig:rmse_per_level}
\end{figure}

General aspects of weight sharing for unrolled algorithms have been extensively discussed in the literature, e.g., see \cite{amj18,hsqdsr19}.
Fig.~\ref{fig:rmse_per_level} gives some insights in the context of our specific approach.
It clearly indicates that weight sharing also changes the reconstruction dynamic within the neural networks.
Earlier iteration steps of the weight-shared $\ItNet$s are more effective, while the non-weight-shared counterparts draw most of their performance from the later steps.
This suggests a trade-off between increasing the model capacity and the difficulty of optimizing the resulting network, while weight sharing forms a simple remedy; cf.~\citet{hsqdsr19}.
However, we conjecture that an improved training strategy for the non-weight-shared networks might unlock the potential of the early-step UNet-blocks.
This could lead to an even larger performance gap between the final reconstruction accuracy of $\ItNet$s with and without weight sharing.
A systematic study of this aspect is left to future research.

\section{Performance on Real-World Image Data}
\label{sec:lodopab}

\begin{figure*}[h!]
	\centering
	\includegraphics[width=\textwidth]{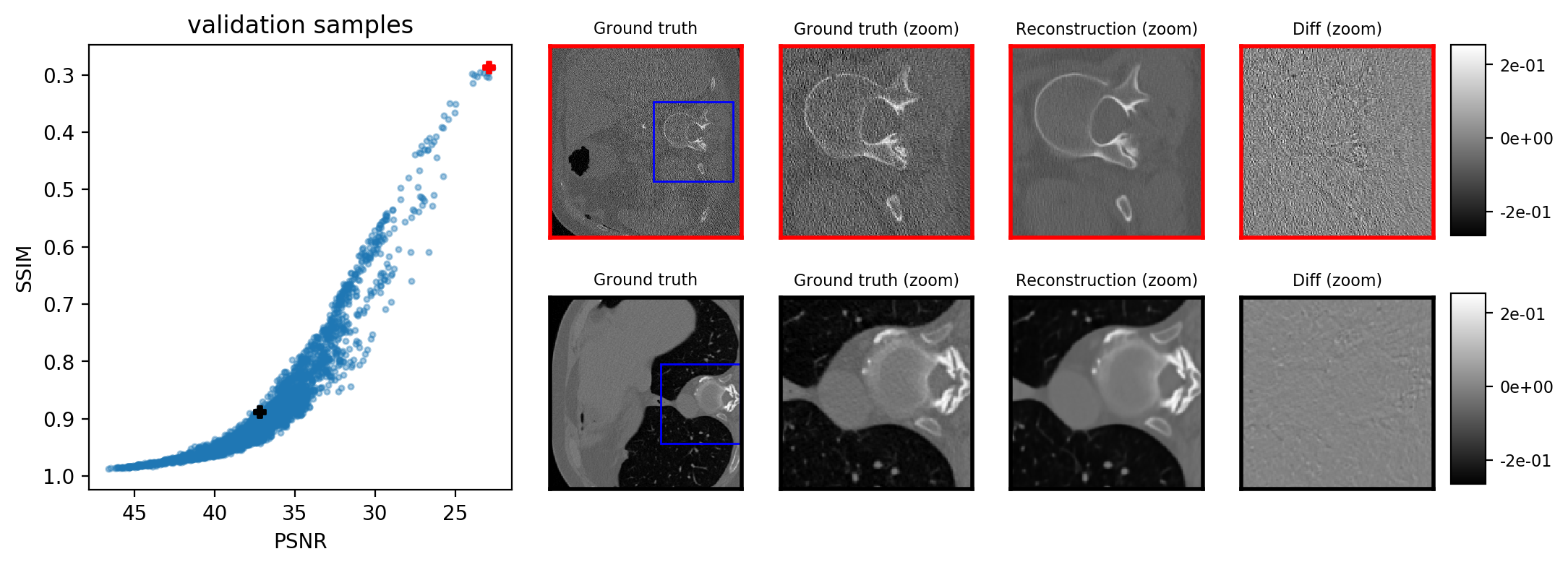}
	\vsp{-.1}
	\caption{\unboldmath\textbf{Results for LoDoPaB CT.} The plot on the left-hand side visualizes the reconstruction performance of the $\ItNet$ with respect to the challenge metrics (SSIM and PSNR) where each blue point corresponds to one image in the LoDoPaB CT validation set (3522 images). We also show individual reconstructions for the red point (worst SSIM) and black point (closest to average SSIM and PNSR) on the right-hand side. Most notable is that the poor SSIM value of the red point is rather due to a low-quality ground-truth image, than a low-quality reconstruction.}
	\label{fig:lodopab_analysis}
\end{figure*}

In order to assess the effectiveness of our method on real-world images and noisy (but still simulated) measurements, we have applied it to the \emph{low-dose parallel beam (LoDoPaB)} CT dataset~\cite{leuschner21}.
This dataset is part of a past challenge and was successfully used to benchmark various deep-learning-based reconstruction schemes.
It consists of 42895 two-dimensional human chest CT slices and their low-intensity measurements, see~\citet{leuschner21} for details on the low-dose setup.
For our case study, we have applied Step~2 and Step~3 of the methodology in Section~\ref{sec:methodology}.\footnote{We have trained an ensemble of five $\ItNet_3$ with initial learning rate of $8\cdot 10^{-5}$ and batch size $2$. As loss function, a combination of the MSE and SSIM was used.}
The resulting $\ItNet$ has reached the first place in the public leaderboard (still open for submissions),\footnote{Team-name: \texttt{RobustAndStable}; public leaderboard on \url{https://lodopab.grand-challenge.org} (accessed on June~7, 2022).} thereby outperforming various other methods, such as the learned primal-dual algorithm~\cite{ao18} (cf.~Table~\ref{tab:res}).
A brief analysis and visualization of our reconstructions results can be found in Fig.~\ref{fig:lodopab_analysis}.
Overall, we conclude that our solution strategy can also achieve state-of-the-art performance on natural image data.

\end{document}
